\documentclass[manuscript]{aastex}

\usepackage{amsmath}

\def\vec#1{\mbox{\boldmath $#1$}}

\slugcomment{Not to appear in Nonlearned J., 45.}

\shorttitle{Capture of planetesimals by waning circumplanetary disks}
\shortauthors{Suetsugu \& Ohtsuki}

\begin{document}


\title{Capture of Planetesimals by Waning Circumplanetary Gas Disks}


\author{Ryo Suetsugu\altaffilmark{1,2}, Keiji Ohtsuki\altaffilmark{2}}
\affil{1.Organization of Advanced Science and Technology, Kobe University, Kobe 657-8501, Japan}
\affil{2. Department of Planetology, Kobe University, Kobe 657-8501, Japan}

\email{suetsugu@buffalo.kobe-u.ac.jp, ohtsuki@tiger.kobe-u.ac.jp}



\begin{abstract}

When gas giant protoplanets grow sufficiently massive, circumplanetary disks would form. 
While solid bodies captured by the circumplanetary disks likely contribute to the growth of the planets and regular satellites around them, some of captured bodies would remain in planet-centered orbits after the dispersal of the disk.
We examine capture and subsequent orbital evolution of planetesimals in waning circumplanetary gas disks using three-body orbital integration. 
We find that some of captured planetesimals can survive in the circumplanetary disk for a long period of time under such weak gas drag. 
Captured planetesimals have semi-major axes smaller than about one third of the planet's Hill radius.
Distributions of their eccentricities and inclinations after disk dispersal depend on the strength of gas drag and the timescale of disk dispersal, and initially strong gas drag and quick disk dispersal facilitates capture and survival of planetesimals.
However, in such a case, final orbital eccentricities and inclinations of captured bodies remain rather large.
Although our results suggest that some of the present irregular satellites of gas giant planets with small semi-major axes would have been captured by gas drag, other mechanisms are required to fully explain their current orbital characteristics.

\end{abstract}





\section{INTRODUCTION}
\label{sec:intro}

Regular satellites of the giant planets in the Solar system are moving on nearly circular and coplanar orbits, thus they are thought to be formed in circumplanetary gas disks.
Solid materials in the circumplanetary disk that formed satellites are supplied from the protoplanetary disk. 
\citet{CW02} assumed that the major building blocks of regular satellites are meter-sized or smaller bodies that are brought to the disk with the gas inflow from the protoplanetary disk. 
On the other hand, supply of solid bodies to circumplanetary disks has been recently studied in detail using orbital integration. 
Assuming an axisymmetric structure for the circumplanetary disk, \citet{F13} performed three-body orbital integrations and examined capture of planetesimals from their heliocentric orbits by the circumplanetary disk. 
They found that planetesimals approaching the circumplanetary disk in the retrograde direction (i.e., in the direction opposite to the motion of the gas) are more easily captured by gas drag than those in the prograde direction (those moving in the same direction as the gas in the disk), because of the larger velocity relative to the gas. 
They also obtained analytically radial distance from the planet where planetesimals with a given size become captured by gas drag.
\citet{T14} examined capture of solid bodies using results of hydrodynamic simulations of gas flow around a growing giant planet and three-body orbital integration for initially circular, non-inclined orbits, and found that accretion efficiency peaks around 10 meter-sized materials. 
These works showed that bodies that are sufficiently large to be decoupled from the gas flow can contribute to the formation of regular satellites.

Influence of captured solid bodies on satellite system formation would vary depending on the timing of capture. 
When planetesimals are captured by gas drag from the circumplanetary disk in the midst of accretion of regular satellites, part of captured planetesimals would contribute to the growth of satellites, while the rest spirals into the central planet \citep{F13, T14, SO16}. 
However, the circumplanetary disk dissipates at some point due to either gap formation in the protoplanetary disk or global dispersal of the protoplanetary disk.
Planetesimals captured by such a waning circumplanetary gas disk would survive in the disk for a long period of time, and may become irregular satellites after the dispersal of the disk.
Also, some of the captured planetesimals would collide with regular satellites during or after the disk dispersal and may influence their surfaces \citep{B10, B13}.

However, capture of planetesimals by weak gas drag from waning circumplanetary disks has not been examined in detail.
\citet{CB04} examined capture of irregular satellites by waning disks in the late stage of planet formation, and discussed the origin of a cluster of prograde irregular satellites of Jupiter. 
Assuming that the cluster members are collisional fragments derived from a single body, they integrated orbits of the cluster progenitor backward in time until it escaped from the planet's Hill sphere, taking account of weak gas drag from the circumjovian disk. 
They found that some planetesimals captured into prograde orbits about Jupiter likely experienced a period of temporary capture before permanent capture.
However, \citet{CB04} mainly focused on the capture of prograde irregular satellites and did not examine capture and orbital evolution of retrograde irregular satellites. 
Also, because of their backward integration, orbital evolution from various initial heliocentric orbits was not examined, and capture rates were not obtained.

In the present work, we examine capture of planetesimals in waning circumplanetary gas disks using three-body orbital integration. 
In addition to the process of capture, we also examine subsequent orbital evolution of captured planetesimals.
We show that some of captured planetesimals can survive in the circumplanetary disk for a long period of time under such weak gas drag. 
Based on results of our orbital integration, we examine relationship between planetocentric orbital elements of captured planetesimals and disk parameters, such as the strength of gas drag or time scale of disk dispersal.
In Section \ref{sec:method}, we describe basic equations, disk model, and numerical methods used in the present work. 
Numerical results on the rates of permanent capture of planetesimals by gas drag are presented in Section \ref{sec:rates}.
In Section \ref{sec:orbit}, we show examples of orbital evolution of planetesimals captured by weak gas drag, and also examine characteristics of long-lived capture orbits in the circumplanetary disk.
In Section \ref{sec:rayleigh}, taking account of gradual dispersal of the circumplanetary disk, we examine distribution of planetocentric orbits of captured planetesimals, and compare them with observed orbital elements of irregular satellites of the giant planets in the Solar System. 
Section \ref{sec:sum} summarizes our results.

\section{THE MODEL AND NUMERICAL METHODS}
\label{sec:method}
\subsection{Basic Equations}

We consider the three-body problem for the Sun, a planet ($M$), and a planetesimal ($m_{\rm s}$), and assume that the planet is on a non-inclined, circular orbit and has a circumplanetary gas disk.
Planetesimals are assumed to be initially on heliocentric orbits with uniform radial distribution in the protoplanetary disk.
We use a rotating coordinate system centered on the planet, where the $x$-axis points radially outward, the $y$-axis points the direction of orbital motion, and the $z$-axis is normal to the $x$-$y$ plane.
Then, the relative motion between the planet and planetesimals can be described by Hill's equation.
We scale time by $\Omega^{-1}$ ($\Omega$ is the planet's orbital angular frequency) and distance by the mutual Hill radius $R_{\rm H}= a_{0}h_{\rm H}$ 
($a_{0}$ is the semi-major axis of the planet), where $h_{\rm H} =\left(\left(M+m_{\rm s}\right)/3M_{\sun}\right)^{1/3}$. 
The non-dimensional equation for the relative motion between the planet and a planetesimal can be written as 
\citep[e.g.,][]{O12, F13}
\begin{equation}
\begin{split}
\ddot{\tilde{x}}&=2\dot{\tilde{y}}+3\tilde{x}-\frac{3\tilde{x}}{\tilde{R}^{3}}+\tilde{a}_{{\rm drag}, x}, \\
\ddot{\tilde{y}}&=-2\dot{\tilde{x}}-\frac{3\tilde{y}}{\tilde{R}^{3}}+\tilde{a}_{{\rm drag}, y}, \\
\ddot{\tilde{z}}&=-\tilde{z}-\frac{3\tilde{z}}{\tilde{R}^{3}}+\tilde{a}_{{\rm drag}, z}, 
\label{eq:hillgas}
\end{split}
\end{equation}
where $\tilde{R}=\sqrt{\tilde{x}^{2}+\tilde{y}^{2}+\tilde{z}^{2}}$ is the normalized distance between the centers of the planet and the planetesimal.
$\vec{\tilde{a}}_{\rm drag}=\vec{a}_{\rm drag}/(R_{\rm H}\Omega^{2})$ is the non-dimensional acceleration due to gas drag, where $\vec{a}_{\rm drag}\equiv\vec{F}_{\rm drag}/m_{\rm s}$ and the gas drag force $\vec{F}_{\rm darg}$ is given by
\begin{eqnarray}
\vec{F}_{\rm drag}=-\frac{1}{2}C_{\rm D}\pi r^{2}_{\rm s}\rho_{\rm gas}u\vec{u}.
\label{eq:gasforce}
\end{eqnarray}
In the above, $C_{\rm D}$ is the drag coefficient (we assume $C_{\rm D}=1$), $r_{\rm s}$ is the radius of the planetesimal, $\rho_{\rm gas}$ is the gas density, and $\vec{u}$ is the velocity of the planetesimal relative to the gas ($u=|\vec{u}|$). 
Using Equation~(\ref{eq:gasforce}), $\vec{\tilde{a}}_{\rm drag}$ can be written as 
\begin{eqnarray}
\vec{\tilde{a}}_{\rm drag}=-\frac{3}{8}C_{\rm D}\frac{\rho_{\rm gas}}{\tilde{r}_{\rm s}\rho_{\rm s}}\tilde{u}\vec{\tilde{u}},
\label{eq:nga}
\end{eqnarray}
where $\rho_{\rm s}$ is the internal density of planetesimals.
When the gas drag can be neglected, Equation~(\ref{eq:hillgas}) holds an energy integral given as \citep[e.g.,][]{NA89, O12}
\begin{eqnarray}
\tilde{E}&=&\frac{1}{2}\left(\dot{\tilde{x}}^{2}+\dot{\tilde{y}}^{2}+\dot{\tilde{z}}^{2}\right)+\tilde{U}(\tilde{x}, \tilde{y}, \tilde{z}),
\label{eq:enegas} 
\end{eqnarray}
where
\begin{eqnarray}
         \tilde{U}(\tilde{x}, \tilde{y}, \tilde{z})=-\frac{1}{2}\left(3\tilde{x}^{2}-\tilde{z}^{2}\right)-\frac{3}{\tilde{R}}+\frac{9}{2}.
\label{eq:potgas}
\end{eqnarray}

\subsection{Disk Structure and Gas Drag Parameter}
Gas accretion flow onto circumplanetary disks shows complicated behavior due to the effects of the planet's gravity, tidal force, and Coriolis force \citep[e.g.,][]{M08}.
However, in the case of large planetesimals that are decoupled from the gas flow, 
the effect of gas drag becomes significant only in the dense part of the disk
in the vicinity of the planet, where the disk can be approximated to be axisymmetric.
Thus, in the present work, we assume an axisymmetric thin circumplanetary disk \citep{F13}.
The radial distribution of the gas density is assumed to be given by a power law, and its vertical structure is assumed to be isothermal.
Under these assumptions, the gas density can be written as 
\begin{eqnarray}	
\rho_{\rm gas}=\frac{\Sigma}{\sqrt{2\pi}h}{\rm exp}\left( -\frac{z^{2}}{2h^{2}}\right),
\label{eq:rhodis}
\end{eqnarray}
where $h=c_{\rm s}/\Omega_{\rm p}$ is the scale height of the circumplanetary disk ($\Omega_{\rm p}$ is the Keplerian orbital frequency around the planet), and  
\begin{eqnarray}
\Sigma = \Sigma_{\rm d}\left(\frac{r}{r_{\rm d}}\right)^{-p}, \;\;\;\;\; c_{\rm s}=c_{\rm d}\left(\frac{r}{r_{\rm d}}\right)^{-q/2}
\label{eq:snum_svelo}
\end{eqnarray}
are the gas surface density and sound velocity, respectively, with $r=\sqrt{x^{2}+y^{2}}$ being the horizontal distance from the planet in the mid-plane.
In the above, $r_{\rm d}=dR_{\rm H}$ is a typical length scale roughly corresponding to the effective size of the circumplanetary disk, 
and $\Sigma_{\rm d}$ and $c_{\rm d}$ are the surface density and sound velocity at $r=r_{\rm d}$ \citep{F13}. 
In our calculations, we set $d=0.2$ and $p=3/2$ based on results of hydrodynamic simulations \citep{M08, T12}, and also assume $q=1/2$ as a simple model \citep{F13}. 
In order to avoid effects of artificial cutoff at $r=r_{\rm d}$, we turn on gas drag when planetesimals enter the planet's Hill sphere. 
Because the gas density decreases rapidly with increasing distance from the planet, this assumption does not affect results of our calculations.

Gas elements in the disk rotate in circular orbits around the planet with velocity slightly lower than the Keplerian velocity ($v_{\rm pK}$) due to radial pressure gradient, as
\begin{eqnarray}
v_{\rm gas}=(1-\eta)v_{\rm pK}.
\label{eq:gasvelo}
\end{eqnarray}
Using Equations (\ref{eq:rhodis}) and (\ref{eq:snum_svelo}), $\eta$ can be written as \citep{T02}
\begin{eqnarray}
\eta = \frac{1}{2}\frac{h^{2}}{r^{2}}\left(p+\frac{q+3}{2}+\frac{q}{2}\frac{z^{2}}{h^{2}}\right).
\end{eqnarray}

When the gas density is given by Equation (\ref{eq:rhodis}), Equation (\ref{eq:nga}) can be rewritten as \citep{F13}
\begin{eqnarray}
\vec{\tilde{a}}_{\rm drag}=-\zeta\tilde{r}^{-\gamma}{\rm exp}\left(-\frac{\tilde{z}^{2}}{2\tilde{h}^{2}}\right)\tilde{u}\vec{\tilde{u}},
\end{eqnarray}
where $h=h_{\rm d}\left(r/r_{\rm d}\right)^{(3-q)/2}$ with $h_{\rm d}$ being the scale height at $r=r_{\rm d}$, $\gamma\equiv p+(3-q)/2$ ($=11/4$ in our model), and $\zeta$ is the non-dimensional parameter representing the strength of gas drag defined by
\begin{eqnarray}
\zeta&\equiv&\frac{3}{8\sqrt{2\pi}}\frac{C_{\rm D}}{r_{\rm s}\rho_{\rm s}}\frac{\Sigma_{\rm d}}{\tilde{h}_{\rm d}}d^{\gamma} \\
      &=& 3 \times 10^{-7} C_{\rm D} \left( \frac{r_{\rm s}}{1 {\rm km}} \right)^{-1} \left( \frac{\rho_{\rm s}}{1 {\rm g\, cm^{-3}}} \right)^{-1} \left( \frac{\Sigma_{\rm d}}{1 {\rm g\, cm^{-2}}} \right) \left( \frac{\tilde{h}_{\rm d}}{0.06} \right)^{-1} \left( \frac{d}{0.2} \right)^\gamma. \nonumber
      \label{eq:zeta_dif}
\end{eqnarray}
We set $\tilde{h}_{\rm d}\equiv h_{\rm d}/R_{\rm H}=0.06$ in the present work \citep{T12, F13}.
For the above fiducial values of $C_{\rm D}$, $d$, and $\tilde{h}_{\rm d}$, we have
\begin{eqnarray}
r_{\rm s}\simeq 0.03\left(\frac{\Sigma_{\rm d}}{1 {\rm g\, cm}^{-2}}\right)\left(\frac{\rho_{\rm s}}{1 {\rm g\, cm}^{-3}}\right) ^{-1}\zeta^{-1}\;\;\;\;\; {\rm cm}.
\end{eqnarray}

\begin{figure}[htbp]
\begin{center}
\plottwo{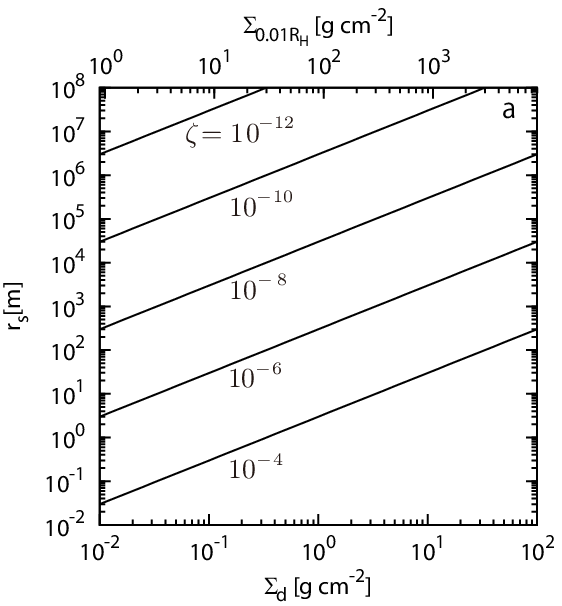}{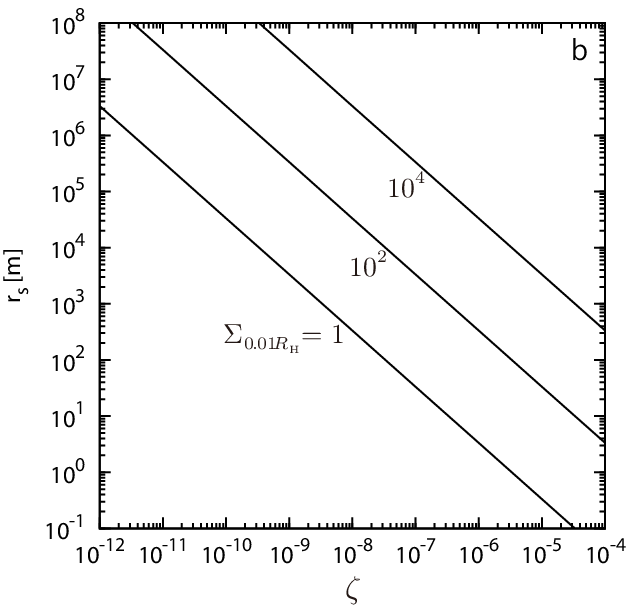}
\end{center}
 \caption{(a) Relationship between planetesimal radius ($r_{\rm s}$) and the gas surface density ($\Sigma_{\rm d}$) at the outer edge ($r=r_{\rm d}$) of a circumplanetary disk for a given value of the non-dimensional gas drag parameter $\zeta$.
Lower and upper horizontal axes show the gas surface density at $r=r_{\rm d}$ (effective size of the circumplanetary disk) and $r=10^{-2}R_{\rm H}$ (roughly corresponding to the radial location of principal regular satellites), respectively. 
The numbers show the assumed values of $\zeta$ for each case ($C_{\rm D}=1$ and $\rho_{\rm s}=1$gcm$^{-3}$).
(b) $r_{\rm s}$ as a function of $\zeta$ for $\Sigma_{0.01R_{\rm H}}=1$, $10^{2}$, and $10^{4}$ gcm$^{-2}$.}
 \label{fig:sur_r}
\end{figure}

Figure \ref{fig:sur_r} shows the relationship between planetesimal radius and gas surface density in a circumplanetary disk for several values of $\zeta$.
In the present work, we consider planetesimals that are large enough to be decoupled from the inflowing gas, with $\zeta \ll 1$.
\citet{F13} used relatively large gas drag parameter $(3\times10^{-9}\leq \zeta\leq10^{-4})$ to examine the contribution of planetesimals to the formation of regular satellites.
If we assume the gas surface density based on the gas-starved disk model \citep[$\Sigma =100\mbox{g cm}^{-2}$ at $r=20R_{\rm J}$;][]{CW02,OI12}, 
the above values of $\zeta$ roughly corresponds to planetesimals with size of $\sim1$m$-100$km.  
In the present work, we investigate capture process of planetesimals by weak gas drag from wanning circumplanetary disks near the last stage of satellite formation. 
Therefore, we will adopt smaller values for the gas drag parameter $(10^{-12}\lesssim \zeta \lesssim10^{-9})$. 
In this case, $\zeta$ roughly corresponds to $r_{\rm s} \sim0.1$km$-1000$km when the surface gas density is one hundredth of the typical value of the gas-starved disk model.
It should be noted that the functional form of $\zeta$ is similar to the inverse of the Stokes number at $r=r_{\rm d}$ \citep{F13}.

\subsection{Orbital Elements of Planet-Centered Orbits}
\label{subsec:twoorb_ircap}
Once planetesimals are captured by the planet's gravity, it is convenient to express their planet-centered orbits using orbital elements based on the two-body problem for the planet and a planetesimal,
although the elements are not constant due to the effect of the solar gravity. 
In the following, the semi-major axis, eccentricity, inclination, and the two-body energy of the planet-centered orbit of a planetesimal are denoted by $a_{\rm p}$, $e_{\rm p}$,  $i_{\rm p}$ and $E_{\rm 2b}$, respectively; and $\tilde{a}_{\rm p}\equiv a_{\rm p}/R_{\rm H}$ and $\tilde{E}_{\rm 2b} \equiv E_{\rm 2b}/(R_{\rm H}\Omega)^2$.
Typically, capture of planetesimals due to gas drag from a circumplanetary disk takes place when they pass through the dense part of the disk in the vicinity of the planet. 
In such a case, we can roughly estimate an upper limit of $\tilde{a}_{\rm p}$ for captured planetesimals by analytic calculation neglecting effects of the tidal potential and the rotation of the coordinate system \citep{SO16}. 
Because the non-dimensional two-body energy can be written as $\tilde{E}_{\rm 2b} = -3/(2\tilde{a}_{\rm p})$, the energy in the three-body problem given by Equation (\ref{eq:enegas}) can be written as
\begin{eqnarray}
\tilde{E}=-\frac{3}{2\tilde{a}_{\rm p}}-\frac{1}{2}\left(3\tilde{x}^{2}-\tilde{z}^{2}\right)+\frac{9}{2},
\label{eq:ene2d3d}
\end{eqnarray}
and planetesimals become permanently captured if $\tilde{E}<0$.
When the tidal potential can be neglected, this can be rewritten in terms of $\tilde{a}_{\rm p}$ as 
\begin{eqnarray}
\tilde{a}_{\rm p} \lesssim 1/3.
\label{eq:acri}
\end{eqnarray}
The above relation shows that captured planetesimals have $a_{\rm p} \lesssim R_{\rm H}/3$ when capture takes place due to energy dissipation in the vicinity of the planet, such as gas drag from the circumplanetary disk.
Although the effect of the tidal potential is important even within the planet's Hill sphere when $\tilde{a}_{\rm p}$ is large and the relation (\ref{eq:acri}) is only an approximate one, 
our numerical results presented below seem to be well explained by this relation.


\subsection{Numerical Method}
\label{subsec:SO14_nu_three}

Orbital evolution of planetesimals captured by gas drag can be divided into two stages \citep[e.g.,][]{CB04}. 
The first stage is temporary capture by the planet, where planetesimals orbit the planet under its gravity but are not yet gravitationally bound within the planet's Hill sphere \citep{IO07, S11, SO13}.
The second stage is orbital decay due to gas drag after they become gravitationally bound within the Hill sphere. 
When gas drag is strong, the first stage is short or even does not exist. 
However, in the case of weak gas drag, the above two-stage evolution is important, as we will show below. 
We define the duration of temporary capture $T_{\rm tc}$ by the time interval between a planetesimal's first passage of the $\tilde{x}$-axis and the time when $\tilde{E}$ becomes negative.

We integrate a large number of orbits by numerically solving Equation~(\ref{eq:hillgas}) with the eighth-order Runge-Kutta integrator (see \citet{S11} and \citet{F13} for details of orbital calculation).
Initially, planetesimals are uniformly distributed radially, and in the case where they initially have non-zero orbital eccentricities ($e$) or inclinations ($i$), 
their initial horizontal  and vertical phase angles ($\tau, \omega$) are also uniformly distributed.
In the following, we will use the scaled eccentricity and inclination defined as $e_{\rm H}=e/h_{\rm H}$ and $i_{\rm H}=i/h_{\rm H}$.
The initial azimuthal distance of the guiding center is set to $\tilde{y}_{0}={\rm max}(100, 20e_{\rm H})$, which is large enough to neglect mutual gravity between the planet and the planetesimal.
In order to evaluate rates of capture with high accuracy,
we divide our numerical simulation into two steps \citep{O93, OI98, S11, F13}. 
In the first-step calculation, initial orbital elements are given with relatively coarse grids with respect to the difference in the semi-major axes of the planet and the planetesimal and the phase angles, and we search for orbits entering within a critical distance $\tilde{r}_{\rm crit}={\rm max}(3, e_{\rm H})$ from the planet.
In the second step, we set finer grids in the vicinity of orbits found in the first-step calculation, and perform orbital integration to evaluate capture rates and other quantities of captured planetesimals.
Orbital integration in the second-step calculations is terminated when one of the following three conditions is met: 
(a) The distance between the planetesimal and the planet becomes large enough again. 
(b) Collision between the planetesimal and the planet is detected; we assume that the physical size of the planet ($R_{\rm p}$) relative to its Hill radius, $\tilde{R}_{\rm p}\equiv R_{\rm p}/R_{\rm H}$, is $10^{-3}$, which corresponds to the physical size of Jupiter.
(c) The energy of the planetesimal becomes negative within the planet's Hill sphere.

From results of orbital calculation, we obtain non-dimensional capture rates per unit surface number density of planetesimals for given $\tilde{r}_{\rm p}$ defined as \citep{S11, F13}    
\begin{eqnarray}
P_{\rm cap}=\int_{}^{}p_{\rm cap}(b_{\rm H}, e_{\rm H}, i_{\rm H}, \tau, \omega)\frac{3}{2}|b_{\rm H}| db_{\rm H}\frac{d\tau d\omega}{(2\pi)^{2}},
\label{eq:gcap}
\end{eqnarray}
where $b_{\rm H}=b/R_{\rm H}=(a-a_{0})/R_{\rm H}$ is the initial semi-major axis of planetesimals relative to the planet scaled by the planet's Hill radius, and
we set $p_{\rm cap}=1$ for captured orbits with $\tilde{E}<0$ and zero otherwise. 
Using $P_{\rm cap}$, the capture rate in a dimensional form is written as $P_{\rm cap}n_{\rm s}R_{\rm H}^{2}\Omega$, 
where $n_{\rm s}$ is the surface number density of planetesimals.

\section{CAPTURE RATES}
\label{sec:rates}

\begin{figure}[htbp]
 \begin{center}
  \includegraphics[width=160mm]{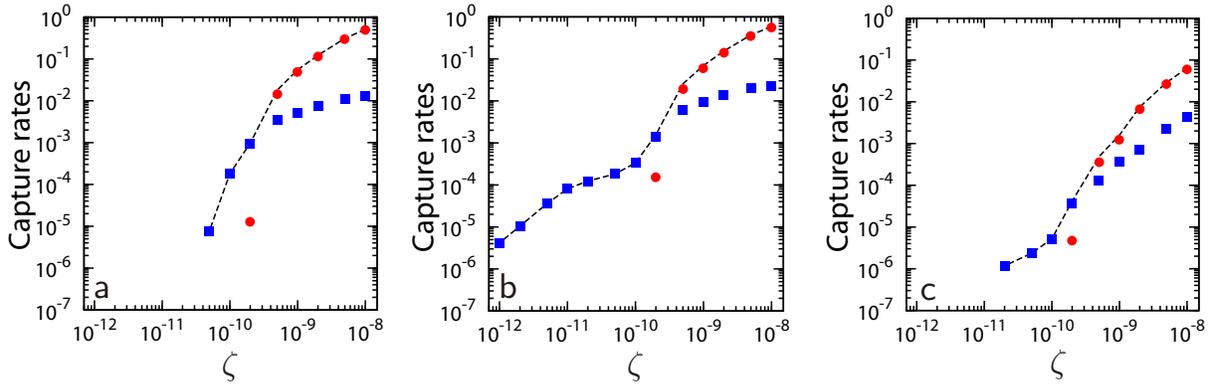}
 \end{center}
 \caption{Rates of permanent capture as a function of $\zeta$ (dashed lines).
 Contributions from captures with $T_{\rm tc}/T_{\rm K}<1$ (red circles) and those with $T_{\rm tc}/T_{\rm K}\geq1$ (blue squares) are also shown.
 (a) $e_{\rm H}=0.1$, (b) $e_{\rm H}=0.7$, and (c) $e_{\rm H}=5$.}
 \label{fig:zera}
\end{figure}

Figure~\ref{fig:zera} shows capture rates in the coplanar case ($i_{\rm H}=0$) as a function of $\zeta$.
Dashed curves represent rates of capture, either in the prograde or retrograde direction.
The blue squares show the case where planetesimals experience temporary capture for more than one orbital period of the planet ($T_{\rm K}$),
while the red circles represent the case of quick permanent capture without such a phase.
Figure~\ref{fig:zera}(a) shows the case of planetesimals with $e_{\rm H}=0.1$.
When $\zeta\gtrsim5\times10^{-10}$, capture rates are dominated by those orbits that did not experience a significant period of temporary capture $(T_{\rm tc}<T_{\rm K})$,
because strong gas drag quickly leads to permanent capture of planetesimals \citep{F13}.
In the case of $\zeta\lesssim2\times10^{-10}$, capture rates with $T_{\rm tc}<T_{\rm K}$ rapidly decrease,
and they become dominated by orbits with a long period of temporary capture.
This behavior can be explained by the transition from single-encounter capture \citep{F13} to multiple-encounter capture with the phase of temporary capture.
Capture rates decrease with decreasing $\zeta$ and eventually disappear when $\zeta<5\times10^{-11}$, because in this case energy dissipation due to gas drag is not sufficient to reduce the relatively high initial energy of nearly circular heliocentric orbits to become long-lived temporary capture orbits.

In the case of $e_{\rm H}=0.7$ (Figure~\ref{fig:zera}(b)), the basic features of the capture rates is similar to the case of $e_{\rm H}=0.1$, and the  transition of capture process (single-encounter or multiple-encounter capture) also appears in a range of the gas drag parameter ($\zeta\simeq2\times10^{-10}-5\times10^{-10}$).
However, in this case even extremely weak gas drag ($\zeta \lesssim 10^{-11}$) leads to permanent capture, because long-lived capture orbits appear for such a value of initial eccentricity ($e_{\rm H}\sim0.7$; see Section \ref{sec:orbit}).
In the dispersion-dominated regime ($e_{\rm H}=5$; Figure~\ref{fig:zera}(c)), the general behavior is similar,  but the capture rates generally decrease owing to the high random velocity \citep{F13}.

Our results show that the phase of temporary capture is necessary for permanent capture to take place when $\zeta \lesssim (2-5) \times10^{-10}$.
This critical value of the gas drag parameter can be explained by the capture radius for single-encounter permanent capture \citep{F13}.
\citet{F13} defined the capture radius $\tilde{R}_{\rm c}$ by the distance from the planet at which the total amount of energy dissipation due to gas drag during a single encounter equals the initial energy of a planetesimal.
In the case of $\zeta= (2-5)\times10^{-10}$, the capture radius for retrograde orbits is given as $\tilde{R}_{\rm c}\simeq (6-8)\times10^{-4}$, which is smaller than the assumed physical size of the planet ($10^{-3}$ in the present case).
This means that even the gas drag in the vicinity of the planet is not strong enough to lead to permanent capture by a single encounter, and multiple encounters with the planet is necessary for capture.

\begin{figure}[htbp]
 \begin{center}
  \includegraphics[width=100mm]{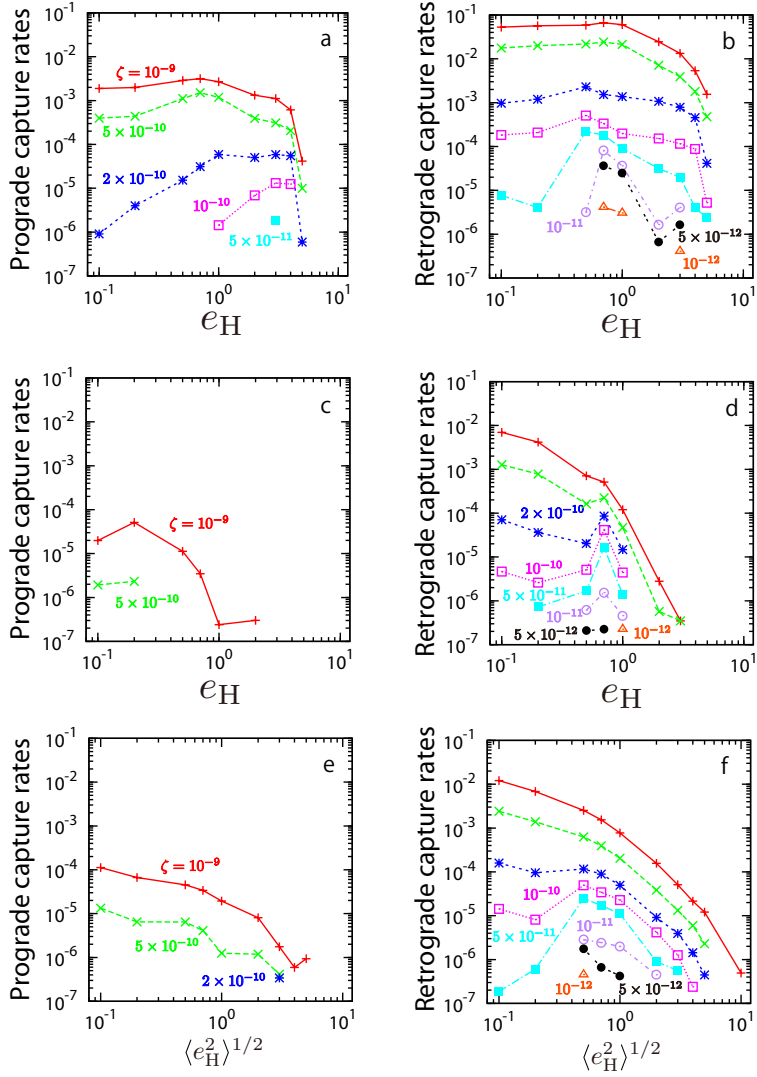}
 \end{center}
 \caption{Rates of permanent capture of planetesimals by the circumplanetary disk, as a function of planetesimals' initial heliocentric orbital eccentricity scaled by $h_{\rm H}$.
Panels (a), (c), and (e) show the rates of capture into prograde orbits, while (b), (d), and (f) show the retrograde case.
Panels (a) and (b) show the coplanar case with $i_{\rm H}=0$, and Panels (c) and (d) represent the case of inclined orbits with $i_{\rm H}=e_{\rm H}/2$.
Panels (e) and (f) show the case with Rayleigh distribution of $e_{\rm H}$ and $i_{\rm H}$ ($\langle e^{2}_{\rm H}\rangle^{1/2}=2\langle i^{2}_{\rm H}\rangle^{1/2}$).}
 \label{fig:caprate}
\end{figure}

Figure~\ref{fig:caprate} shows the plots of the rates of permanent capture for several values of the gas drag parameter, as a function of planetesimals' initial eccentricity. 
Panels (a) and (b) show results in the prograde and retrograde cases for coplanar orbits ($i_{\rm H}=0$), respectively.
In the case of relatively strong gas drag ($\zeta\gtrsim5\times10^{-10}$), capture typically takes place in a single encounter with the planet (Figure~\ref{fig:zera}). 
In this case, the capture rate hardly depends on the eccentricity in the shear-dominated regime (i.e., $e_{\rm H}\lesssim1$), where relative velocity between the planetesimal and the planet is dominated by the Kepler shear, while it monotonically decreases with increasing eccentricity for $e_{\rm H}\gtrsim1$, because large velocity relative to the planet shortens interaction time between the gas and approaching planetesimals.  
On the other hand, in the case of weak gas drag ($\zeta\lesssim10^{-10}$), prograde capture occurs only in a narrow range of $e_{\rm H}$ at $1\lesssim e_{\rm H} \lesssim4$.
This is because long-lived prograde temporary capture occurs only at $e_{\rm H}\simeq3$ \citep{S11, SO13}, and permanent capture via such temporary capture phase is dominant when gas drag is weak.

Figure \ref{fig:caprate}(b) shows the rates of capture in the retrograde direction.
When $\zeta\gtrsim 10^{-10}$, the general behavior is similar to the the prograde case.
Retrograde capture rates are higher than prograde capture rates because of the strong gas drag resulting from the high relative velocity with the gas. 
In the case of weak gas drag ($\zeta\lesssim5\times10^{-10}$), capture rates have a peak at $e_{\rm H}\simeq0.7 - 1$.
This is because planetesimals with relatively low energy ($\tilde{E}\simeq1.5 - 2$) can enter the Hill sphere when $e_{\rm H}\gtrsim0.5$ and are transferred into long-lived capture orbits due to the weak gas drag,
while such long-lived temporary capture orbits with low energy disappear when $e_{\rm H}\geq1$. 
Although \citet{S11} found that long-lived, large temporary capture orbits in the retrograde direction outside of the planet's Hill sphere (called type-E orbits after epicyclic motion) are common for large value of $e_{\rm H}$ $(>3)$, we find that the energy of this type of orbits is too large to become permanently captured with weak gas drag.

Figures \ref{fig:caprate}(c) and (d) show prograde and retrograde capture rates for initially inclined orbits with $i_{\rm H}=e_{\rm H}/2$.
Planetesimals can be captured in the case of strong gas drag ($\zeta=10^{-9}$) and low inclination, 
but capture does not take place with weak gas drag ($\zeta<5\times10^{-10}$), because long-lived capture orbits do not appear in this velocity regime.
When planetesimals' inclinations become large, planetesimals penetrate the disk nearly vertically and suffer from significant gas drag only for a short period of time, 
and even temporary capture is not helpful because the amount of energy dissipation is rather small (Figures~\ref{fig:caprate}(c), \ref{fig:caprate}(d)).
In the case of $\zeta<2\times10^{-10}$, the retrograde capture rates have a peak at $e_{\rm H}\simeq0.7$ as in the coplanar case.
We also examined the case where planetesimals' orbital eccentricities and inclinations have a Rayleigh distribution ($\langle e_{\rm H}^{2}\rangle^{1/2} = 2\langle i_{\rm H}^{2}\rangle^{1/2}$; Figures~\ref{fig:caprate}(e), (f)).
The general behavior is similar to the case of $i_{\rm H}=e_{\rm H}/2$, but capture takes place even for large velocity dispersions ($\langle e_{\rm H}^{2}\rangle^{1/2} \gtrsim 3$) because of contribution from low-velocity orbits.
Retrograde capture rates decrease with increasing eccentricities when $\zeta\gtrsim2\times10^{-10}$ and have a peak at $\langle e^{2}_{\rm H}\rangle^{1/2}\simeq0.5-0.7$ when $\zeta\lesssim2\times10^{-10}$ (Figure~\ref{fig:caprate}(f)), which is similar to the case of $i_{\rm H}=e_{\rm H}/2$ (Figure~\ref{fig:caprate}(d)).


\section{LONG-LIVED CAPTURE ORBITS IN WANING CIRCUMPLANETARY DISKS}
\label{sec:orbit}
\begin{figure}[htbp]
 \begin{center}
  \includegraphics[width=140mm]{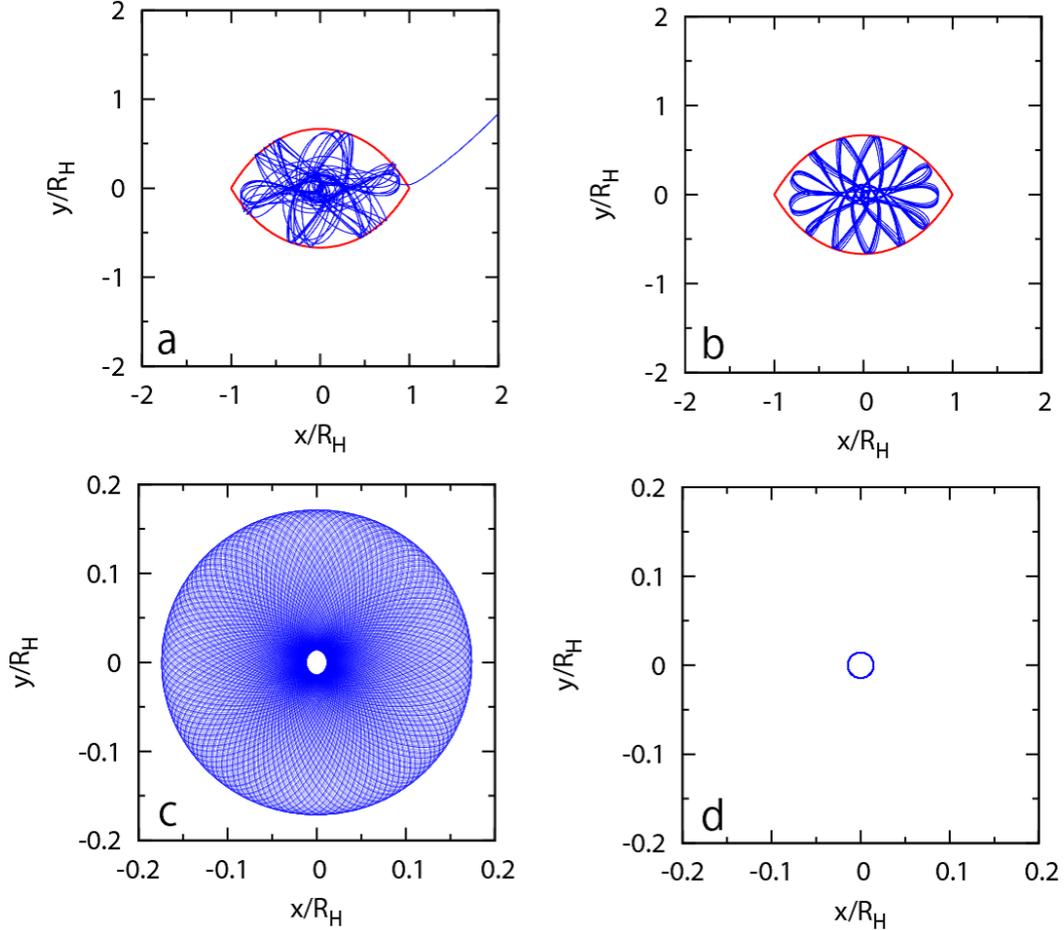}
 \end{center}
 \caption{
Evolution of a long-lived prograde captured orbit ($b_{\rm H}=4.885294$, $e_{\rm H}=3$, $i_{\rm H}=0$, $\tau=0.019716$, $\zeta=10^{-10}$).
Each panel shows orbital behavior for a period of $\sim20T_{\rm K}$ at different stages of the evolution of the same orbit in the rotating coordinate system centered on the planet.
The lemon-shaped curve shows the planet's Hill sphere.
(a) During temporary capture ($t\simeq0$), 
(b) immediately after the planetesimal becomes permanently captured ($t\simeq300$ $T_{\rm K}$), 
(c) at an intermediate stage where the orbit is shrinking due to gas drag ($t\simeq8000$ $T_{\rm K}$), 
and (d) immediately before its semi-major axis becomes smaller than 0.01 times the planet's Hill radius ($t\simeq16000$ $T_{\rm K}$).
}
 \label{fig:exam_pro}
\end{figure}

\begin{figure}[htbp]
 \begin{center}
  \includegraphics[width=150mm]{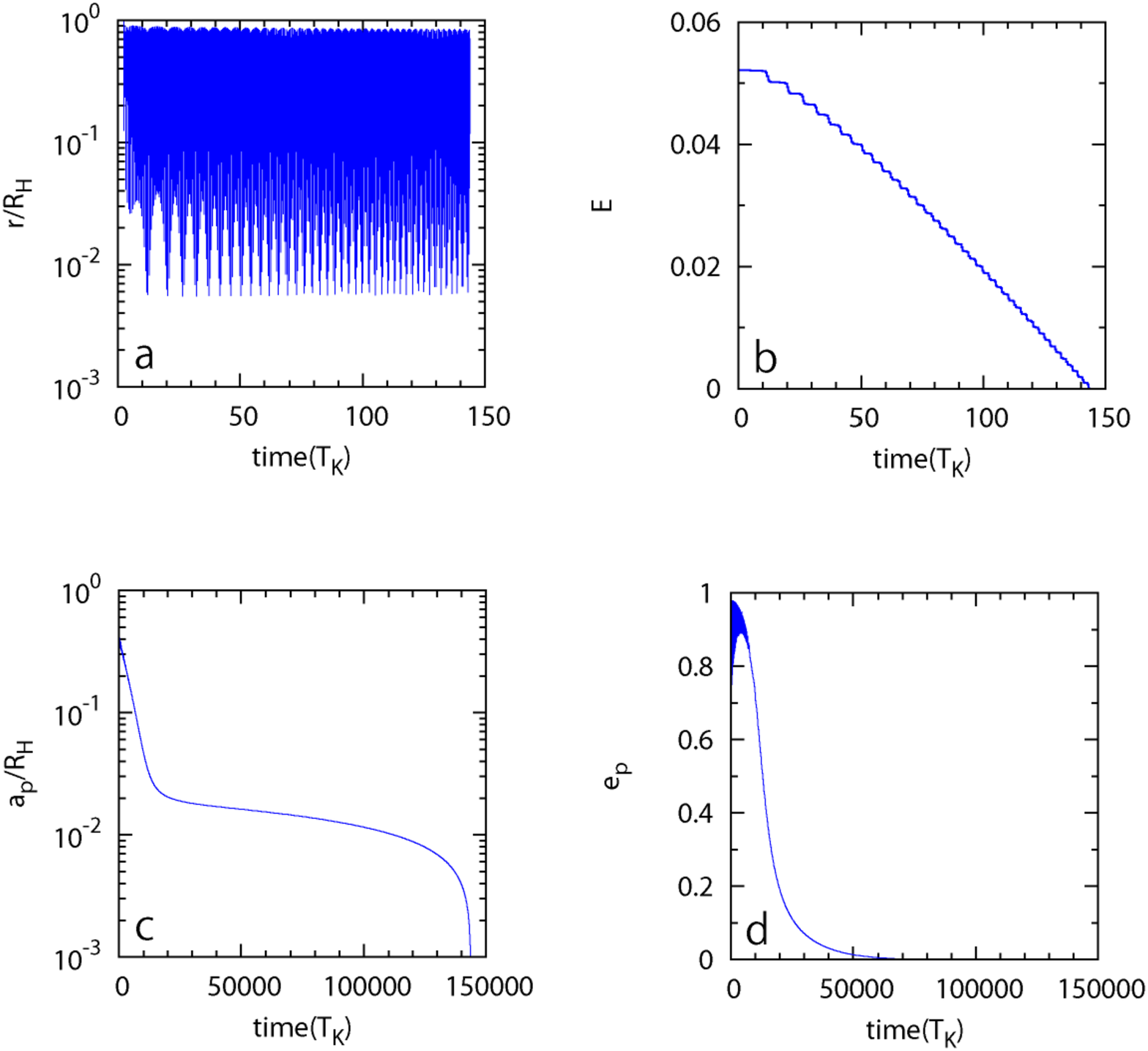}
 \end{center}
 \caption{Change of several quantities for the prograde orbit shown in Figure~\ref{fig:exam_pro}. 
(a) Radial distance from the planet. (b) Energy. (c) Semi-major axis. (d) Eccentricity.}
 \label{fig:enerd_p}
\end{figure}

Here, we show examples of orbits of temporarily captured planetesimals that lead to permanent capture under weak gas drag.
Figure \ref{fig:exam_pro} shows an example of long-lived orbits in the prograde direction, and Figure~\ref{fig:enerd_p} shows changes of various quantities during the evolution.
The four panels in Figure \ref{fig:exam_pro} show orbital behavior at four different phases of evolution of an orbit with initial eccentricity of $e_{\rm H} = 3$.
In the case of orbits with $e_{\rm H} \simeq 3$ in the gas-free environment, planetesimals can enter the planet's Hill sphere through the vicinity of the Lagrangian points ($L_{1}$ or $L_{2}$) and become temporarily captured in the prograde direction for a long time \citep[called type-H orbits in][]{S11, SO13}.
Figure~\ref{fig:exam_pro}(a) shows that such long-lived capture is possible also under weak gas drag.
During this phase of temporary capture, the planetesimal's radial distance from the planet oscillates rapidly, but it remains larger than about 0.005 $R_{\rm H}(=5\tilde{R}_{\rm p})$, and the planetesimal avoids penetrating the dense part of the circumplanetary disk (Figure~\ref{fig:enerd_p}(a)). 
As a result, the planetesimal loses its energy rather slowly, and becomes permanently captured in about $144T_{\rm K}$ (Figure~\ref{fig:enerd_p}(b)). 
Figures \ref{fig:exam_pro}(b), (c), and (d) show snapshots of the orbital evolution after the planetesimal becomes permanently captured.
The semi-major axis of the orbit gradually decreases as the planetesimal loses energy and angular momentum due to gas drag (Figure \ref{fig:enerd_p}(c)). 
The eccentricity $e_{\rm p}$ of the orbit remains rather high ($e_{\rm p}\simeq0.8-1$) immediately after permanent capture (Figure~\ref{fig:enerd_p}(d)).
When the eccentricity becomes as small as 0.2, the rate of orbital decay significantly decreases (Figure~\ref{fig:exam_pro}(d) and Figure~\ref{fig:enerd_p}(c)).
As a result, the lifetime of permanently captured planetesimals in the circumplanetary gas disk can be as long as $10^{5}T_{\rm K}$.

\begin{figure}[htbp]
 \begin{center}
  \includegraphics[width=140mm]{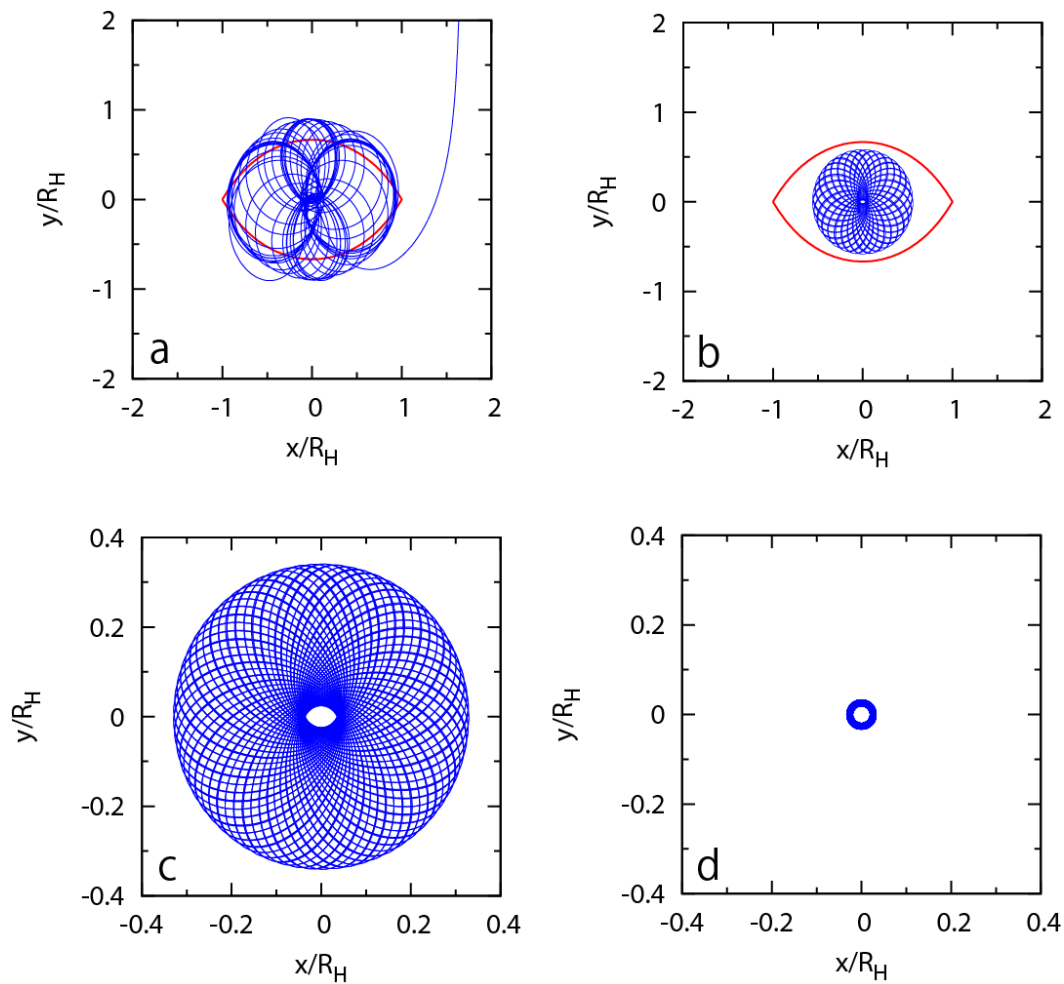}
 \end{center}
 \caption{Same as Figure~\ref{fig:exam_pro}, but the case of capture in the retrograde direction ($b_{\rm H}=2.737524$, $e_{\rm H}=0.7$, $i_{\rm H}=0$, $\tau=5.777929$, $\zeta=10^{-10}$). ~
Orbital evolution for a period of $\sim20T_{\rm K}$ is shown for (a) $t\simeq0$, (b) $t\simeq3232$ $T_{\rm K}$, (c) $t\simeq7070$ $T_{\rm K}$ and (d) $t\simeq14100$ $T_{\rm K}$.}
 \label{fig:exam_ret}
\end{figure}

\begin{figure}[htbp]
 \begin{center}
  \includegraphics[width=110mm]{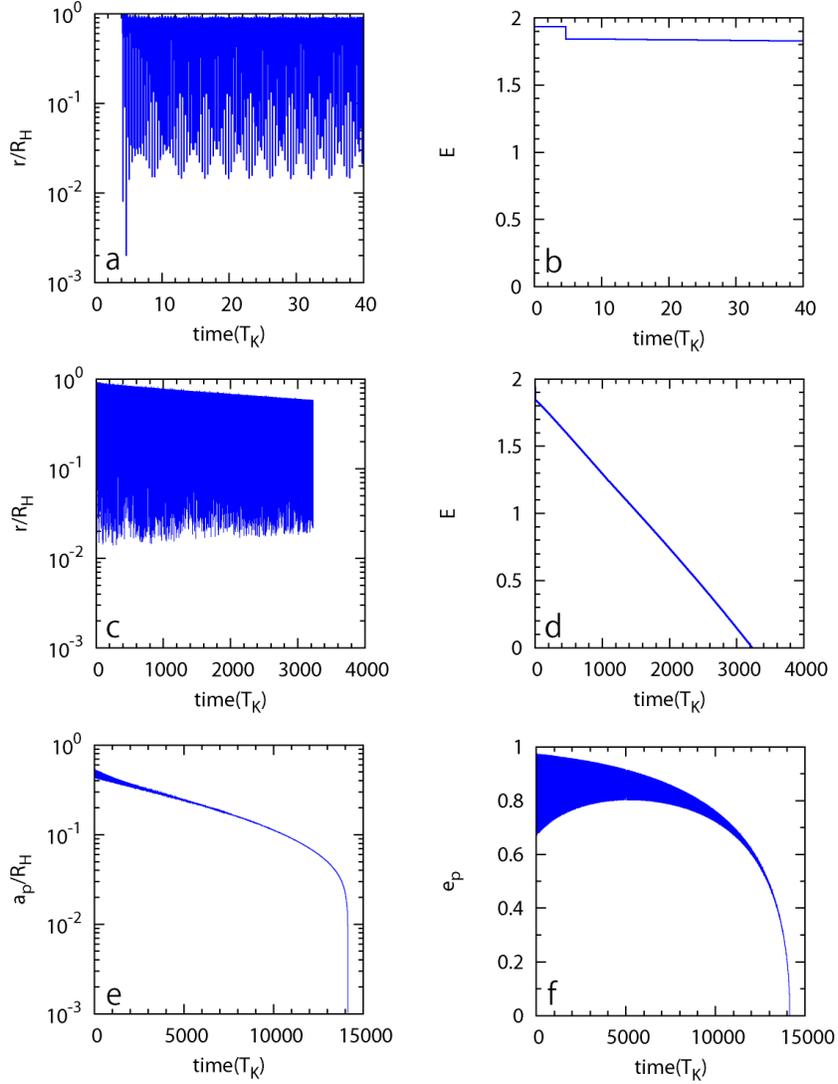}
 \end{center}
 \caption{Changes of radial distance from the planet (Panels (a) and (c)) and energy (Panels (b) and (d)) of a temporarily captured planetesimal in the retrograde direction (the one shown in Figure~\ref{fig:exam_ret}).
Panels (a) and (b) show the initial evolution for $0\leq t\leq 40T_{\rm K}$, and Panels (c) and (d) show the full evolution until the planetesimal becomes permanently captured. 
Bottom panels show changes of semi-major axis (e) and eccentricity (f) of the same orbit.}
 \label{fig:enerd_r}
\end{figure}

Planetesimals captured in the retrograde direction tend to spiral into the planet rather quickly because of their large velocity relative to the gas \citep{F13}. 
However, we find that long-lived retrograde capture orbits exist when planetesimals' initial heliocentric orbit have $e_{\rm H}\simeq0.7-1$ as we mentioned in Section \ref{sec:rates}.
Figure~\ref{fig:exam_ret} shows an example of such long-lived retrograde capture orbits ($e_{\rm H}=0.7$), with Figure~\ref{fig:exam_ret}(a) showing the orbital behavior during the phase of temporary capture.
This type of long-lived temporary capture orbits have energy in the range of  $1.5\lesssim\tilde{E}\lesssim2$ and were not found in the previous study in the gas-free environment \citep{S11}.
In the case of the orbit shown in Figures~\ref{fig:exam_ret} and \ref{fig:enerd_r}, the planetesimal undergoes a close encounter with the planet at $t \simeq 4 T_{\rm K}$ from the beginning of the phase of the temporary capture, and energy dissipation at this encounter transfers the planetesimal into the temporary capture orbit. 
During this phase, its radial distance from the planet remains larger than $0.02R_{\rm H}=20R_{\rm p}$ (Figures~\ref{fig:enerd_r} (a) and (c)). 
As a result, the planetesimal's energy decreases slowly, and the planetesimal becomes permanently captured at $3200T_{\rm K}$ (Figure~\ref{fig:enerd_r}(d)).
After permanently captured, its semi-major axis gradually decreases due to gas drag while eccentricity remains rather high (Figures~\ref{fig:enerd_r}(e) and (f)), 
showing evolution of the orbital shape similar to the prograde case. 
The timescale of the orbital evolution is much longer than the case of strong gas drag \citep{F13}.
In the case shown here, the planetesimal remains in the region with $\tilde{a}_{\rm p}>0.1$ for more than $10^{4}T_{\rm K}$.

\begin{figure}[htbp]
 \begin{center}
  \includegraphics[width=140mm]{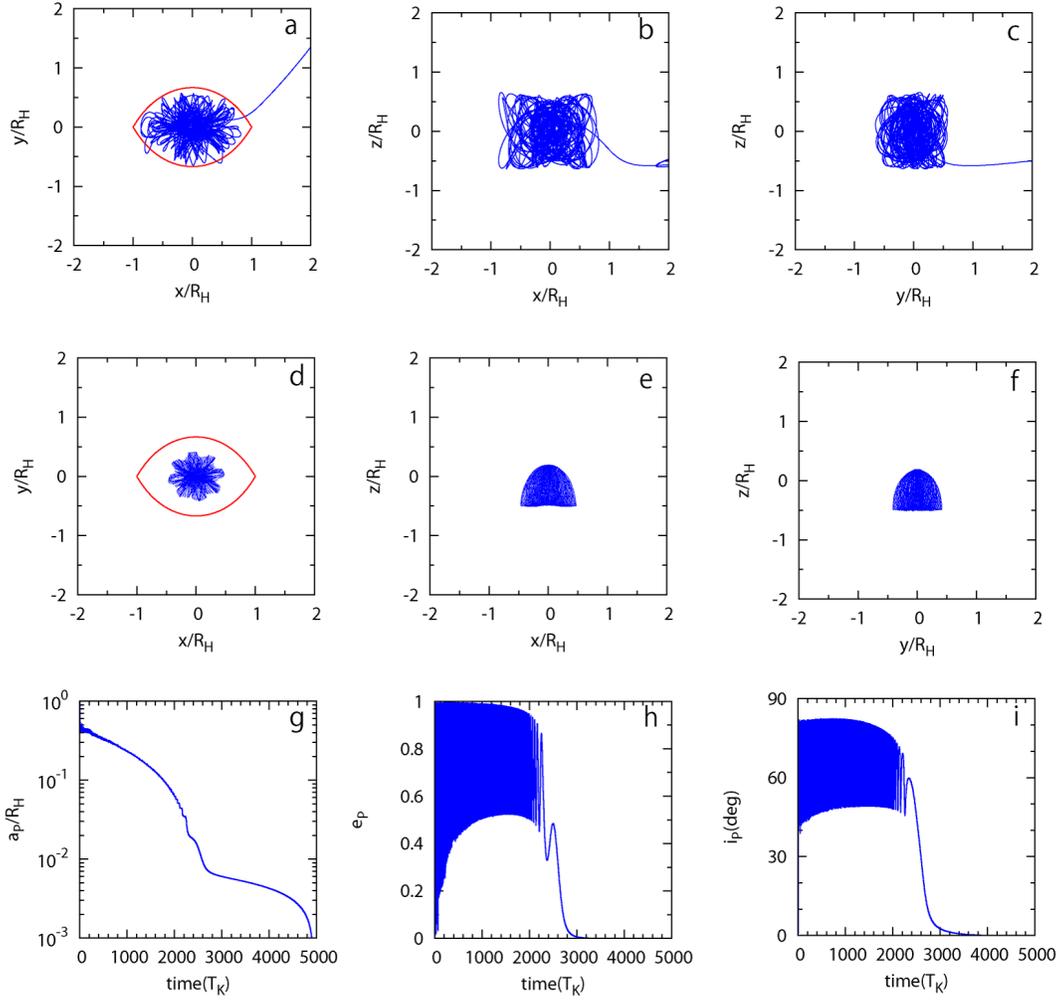}
 \end{center}
 \caption{Evolution of a long-lived prograde captured orbit for the case of initially inclined orbits ($b_{\rm H}=3.91$, $e_{\rm H}=1.917924$, $i_{\rm H}=0.6828253$, $\tau=6.269109$, $\omega=1.324799$, $\zeta=10^{-9}$).
 Each panel shows orbital behavior for a period of $\sim20T_{\rm K}$ at different stages of the evolution of the same orbit in the rotating coordinate system centered on the planet.
Panels (a), (b) and (c) show long-lived capture orbits during temporary capture ($t\simeq0$), viewed from three directions. 
Panels (d), (e) and (f) show orbital behavior immediately after the planetesimal becomes permanently captured ($t\simeq721-741T_{\rm K}$).
Lower panels show changes of semi-major axis (g), eccentricity (h), and inclination (i) of the same orbit for a longer period of time.} 
 \label{fig:orbit_pro3d}
\end{figure}

\begin{figure}[htbp]
 \begin{center}
  \includegraphics[width=140mm]{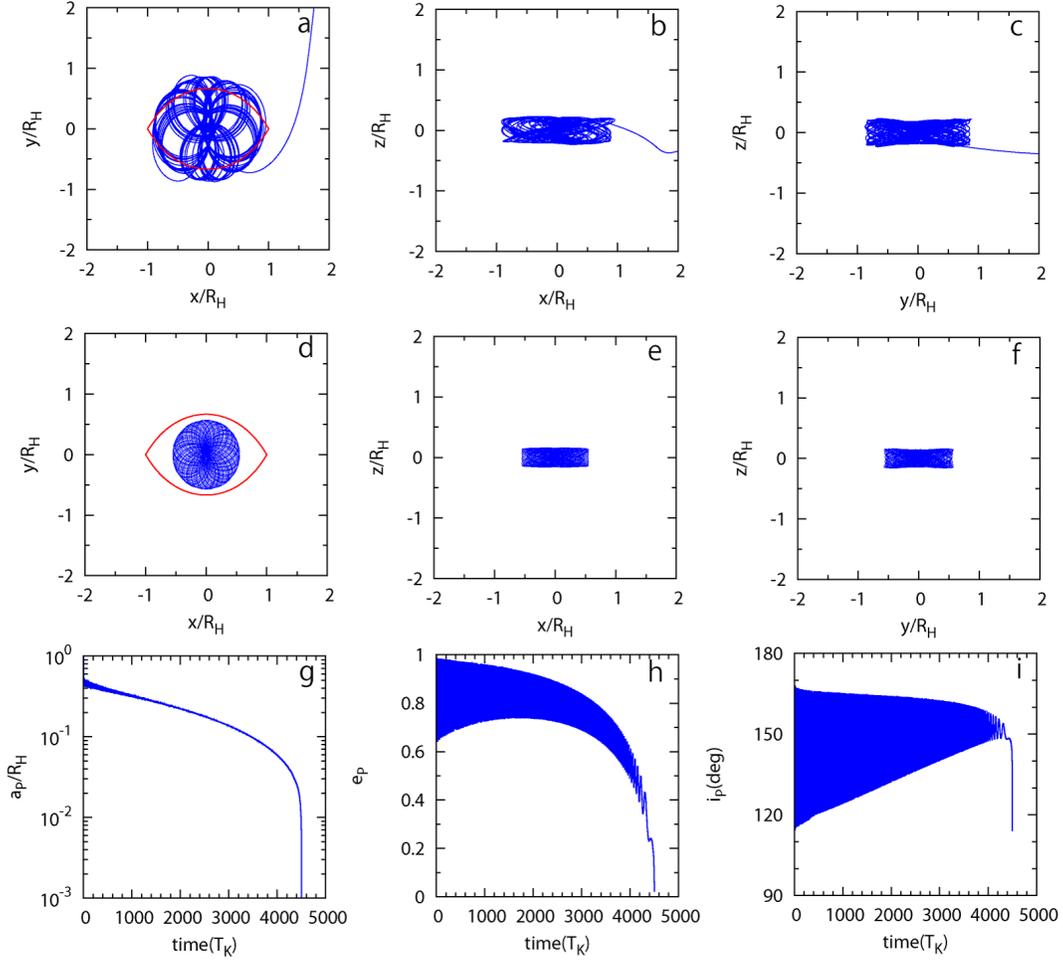}
 \end{center}
 \caption{Same as Figure~\ref{fig:orbit_pro3d}, but the case of capture in the retrograde direction ($b_{\rm H}=2.78$, $e_{\rm H}=0.6198413$, $i_{\rm H}=0.3732988$, $\tau=5.978928$, $\omega=1.156581$, $\zeta=10^{-9}$). 
Orbital evolution for a period of $\sim20T_{\rm K}$ is shown for (a) $t\simeq0T_{\rm K}$, (b) $t\simeq3700T_{\rm K}$, (c) $t\simeq7400T_{\rm K}$ and (d) $t\simeq12600T_{\rm K}$.
Panels (a), (b) and (c) show long-lived capture orbits during temporary capture ($t\simeq0T_{\rm K}$), viewed from three directions. 
Panels (d), (e) and (f) show orbital behavior immediately after the planetesimal becomes permanently captured ($t\simeq1027-1047T_{\rm K}$).
Lower panels show changes of semi-major axis (g), eccentricity (h)  and inclination (i) of the same orbit for a longer period of time.
}
 \label{fig:orbit_ret3d}
\end{figure}

A similar process of permanent capture via the phase of temporary capture can be seen in the case of initially inclined orbits, but the orbital behavior becomes rather complicated (Figures~\ref{fig:orbit_pro3d} and \ref{fig:orbit_ret3d}).
Figure~\ref{fig:orbit_pro3d} shows an example of long-lived orbits in the prograde direction, where a planetesimal undergoes a phase of temporary capture for $T_{\rm tc} \simeq 700T_{\rm K}$  (Figures~\ref{fig:orbit_pro3d}(a), (b), (c)) before permanently captured (Figures~\ref{fig:orbit_pro3d}(d), (e), (f)).
As in the case of $i_{\rm H}=0$, evolution of semi-major axis slows down when the eccentricity becomes $\lesssim0.2$ (Figures~\ref{fig:orbit_pro3d} (g) and (h)), and the inclination decreases rapidly with the rapid decrease of eccentricity (Figure~\ref{fig:orbit_pro3d}(i)). 
Permanent capture in the retrograde direction via the phase of temporary capture can also be seen for initially inclined orbits (Figures~\ref{fig:orbit_ret3d}).
The orbital evolution of captured planetesimals in this case is also similar to the coplanar case.

\section{CAPTURE BY WANING CIRCUMPLANETARY DISK}
\label{sec:rayleigh}

So far we have assumed that the gas drag parameter $\zeta$ is constant during orbital integration for a given set of parameters. However, in the case of capture of planetesimals by a dissipating circumplanetary gas disk, 
the strength of gas drag weakens and $\zeta$ gradually decreases. 
In this section, we examine capture of planetesimals by a waning gas disk, 
and investigate orbital distribution of planetesimals that become captured and survive in the circumplanetary disk.

\subsection{Numerical Methods}

Effects of a waning gas disk on orbital evolution of planetesimals depend on the value of $\zeta$ before disk dispersal and the timescale of the dispersal, 
the latter depending on the process of disk dispersal \citep{S10}. 
When a giant planet becomes massive enough to form a complete gap in the protoplanetary gas disk, 
gas inflow on the planet and the circumplanetary disk is cutoff in a short timescale ($\sim 10^{2}-10^{4}$ years). 
On the other hand, the dispersal of the circumplanetary disk due to the dispersal of the protoplanetary gas disk as a whole takes place on a longer timescale ($\sim 10^{6}$ years). 
Since the mechanism of the dispersal of the circumplanetary gas disk is not fully understood, we adopt a simple model. 
We take account of the disk dispersal by assuming that the gas drag parameter $\zeta$ in our orbital integration is given as a function of time as
\begin{eqnarray}
\zeta (\tilde{t})=\zeta_{\rm ini}{\rm exp}\left(-\tilde{t}/\tau_{\rm dis}\right),
\label{eq:zeta_dis}
\end{eqnarray}
where $\zeta_{\rm ini}$ is the value of $\zeta$ at the time of the beginning of the disk dispersal, 
and $\tau_{\rm dis}$ is the dispersal timescale.
We examine cases with ten different values of $\tau_{\rm dis}$ in a range $\tau_{\rm dis}/T_{\rm K} = 10 - 10^4$, 
where the maximum values of $\tau_{\rm dis}$ has been practically chosen to avoid excessively long computing time. 
As for $\zeta_{\rm ini}$, we examine eight values in a range $\zeta_{\rm ini} = 5 \times 10^{-11}-1\times 10^{-8}$. 
We perform simulations for these eighty combinations of $\zeta_{\rm ini}$ and $\tau_{\rm dis}$, 
and examine the dependence of the distribution of captured planetesimals on these parameters.

Numerical methods are basically the same as those described in the previous sections, but are slightly modified.  
We assume that orbital eccentricities and inclinations follow a Rayleigh distribution with given r.m.s. eccentricities and inclinations 
($\langle e_{\rm H}^2 \rangle^{1/2} = 2 \langle i_{\rm H}^2 \rangle^{1/2}$), and that orbital phase angles are randomly distributed.
In order to account for the supply of incoming planetesimals due to the Kepler shear, 
the number of incoming planetesimals with $b_{\rm H}$ are assumed to be proportional to $b_{\rm H}$. 
As in the simulations presented in the previous sections, 
we adopt two-step simulations to save computing time. 
For each orbit, we define $t = 0$ by the time the planetesimal crosses the $x$-axis; the gas drag parameter $\zeta$ starts decreasing from this moment, 
and we continue orbital integration until $t=4\tau_{\rm dis}$. 
Orbital evolution of captured planetesimals continues even after $t=4\tau_{\rm dis}$, 
but we stop our integration at this point to save computing time. 
At this point, the drag parameter becomes two orders of magnitude smaller than its initial value, 
and we can examine effects of disk dispersal. 
Because of this newly-adopted condition for the termination of orbital integration, 
we do not adopt the condition (c) (i.e., negative energy due to dissipation by gas drag) in the present set of simulations. 
In the following, we focus on the orbital distribution of planetesimals that are permanently captured at the end of the simulation, 
and those that are temporarily captured at that time are not included.

The above numerical method neglects the effect of newly captured planetesimals after the beginning of the dispersal of the circumplanetary disk.
However, we can roughly estimate distribution of such bodies, using results of our orbital integrations.  
The relation of the gas drag parameters for different times ($\tilde{t}_{1}$, $\tilde{t}_{2}(=\tilde{t}_{1}+\tilde{t}')$) can be written as $\zeta(\tilde{t}_2)=\zeta_{\rm ini} {\rm exp}(-(\tilde{t}_{1}+\tilde{t}')/\tau_{\rm dis})=\zeta(\tilde{t}_{1}){\rm exp}(-\tilde{t}'/\tau_{\rm dis})$. 
This indicates that $\zeta(\tilde{t}_{2})$ corresponds to the gas drag parameter with $\zeta_{\rm ini}=\zeta(\tilde{t}_{1})$ and $\tilde{t}=\tilde{t}'$.
Thus, for a given values of $\tau_{\rm dis}$, the distribution of captured planetesimals including the effect of planetesimals captured  after the beginning of the disk dispersal can be obtained by superposition of results for different initial gas drag parameters.
Although we do not include such an effect in the following, some important insights can be obtained from our results, as shown below.

\subsection{Distribution of Orbital Elements of Captured Planetesimals}
\subsubsection{Case of Low Velocity Dispersion}

\begin{figure}[htbp]
 \begin{center}
  \includegraphics[width=165mm]{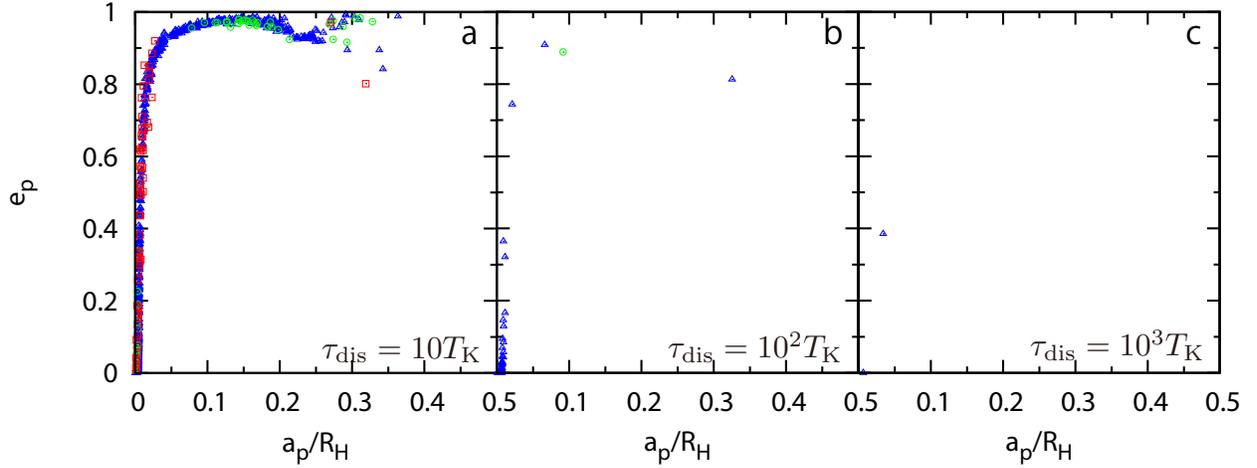}
 \end{center}
 \caption{Distribution of orbital elements of planetesimals captured into prograde orbits at the end of simulation ($\langle e_{\rm H}^{2} \rangle^{1/2}=0.5$ and $\zeta_{\rm ini}=10^{-8}$).
Panels (a), (b) and (c) show the results for $\tau_{\rm dis}=10, 10^{2}$ and $10^{3}T_{\rm K}$, respectively. 
Different marks represent inclinations of planet-centered orbits.
Triangles, circles and squares represent $0^\circ\leq i_{\rm p}<30^\circ$, $30^\circ\leq i_{\rm p}<60^\circ$ and $60^\circ\leq i_{\rm p}<90^\circ$, respectively.
No permanently captured orbits remain in the case of $\tau_{\rm dis}=10^{4}T_{\rm K}$.}
 \label{fig:e0.5pro}
\end{figure}

First, we examine the case where planetesimals are initially on their heliocentric orbits with low velocity dispersion ($\langle e_{\rm H}^2 \rangle^{1/2} =0.5$). 
Figure~\ref{fig:e0.5pro} shows the plots of eccentricities of planet-centered orbits as a function of semi-major axis at the end of simulation, for planetesimals captured into the prograde direction ($\zeta_{\rm ini} = 10^{-8}$). 
Panels (a) to (c) show results for different values of $\tau_{\rm dis}$, and different marks represent the difference in the orbital inclinations of planet-centered orbits. 
When the velocity dispersion of planetesimals on heliocentric orbits is small, 
those planetesimals that enter the planet's Hill sphere tend to have rather high energy \citep{OI98, S11}.
Since there are no long-lived temporary capture orbits in the prograde direction with such a high energy \citep{S11}, 
the lifetime of captured planetesimals in the circumplanetary disk is rather short. 
If the disk gas dissipates in a short timescale, a significant number of captured planetesimals can survive around the planet 
(Figure~\ref{fig:e0.5pro}(a)), but if the disk gas remains for a long time, captured planetesimals spiral into the planet quickly and their survival is difficult (Figures~\ref{fig:e0.5pro}(b), (c)). 
In Figure~\ref{fig:e0.5pro}(a), those captured planetesimals with relatively large $a_{\rm p}$'s tend to have $e_{\rm p} \simeq 1$, while $e_{\rm p}$ decreases rapidly at small $a_{\rm p}$ due to strong gas drag. 
On the other hand, some of the captured planetesimals have rather large orbital inclinations, reflecting their initial orbital inclinations. 

Figure~\ref{fig:e0.5ret} shows numerical results for those captured in the retrograde direction. 
The upper panels show the case with $\zeta_{\rm ini} = 10^{-8}$, 
and the lower panels show the case with weaker gas drag with $\zeta_{\rm ini} = 10^{-10}$. 
The four panels in each row represent cases with different values of $\tau_{\rm dis}$. 
As in the retrograde case, those with small $a_{\rm p}$'s tend to have small eccentricities. 
On the other hand, in the retrograde case, there are long-lived capture orbits even if planetesimals are initially on heliocentric orbits with low velocity dispersion \citep{S11}. 
As a result, a significant number of planetesimals survive even with large values of $\tau_{\rm dis}$ (Figure~\ref{fig:e0.5ret}(b)-(d)). However, with increasing $\tau_{\rm dis}$, orbital eccentricity of captured planetesimals decrease because of the longer duration of interaction time with the gas disk. 
If the gas drag is weak, capture is difficult even in the retrograde direction (Figure~\ref{fig:e0.5ret}(e)-(h)). 
In this case, capture is dominated by temporary capture into long-lived orbits, followed by gradual decrease of energy due to gas drag. 
Permanent capture is difficult if $\tau_{\rm dis}$ is too short, because planetesimals do not have sufficient time for interaction with the gas (Figure~\ref{fig:e0.5ret}(e)). 
Also, the number of surviving planetesimals becomes small when $\tau_{\rm dis}$ is too long, because captured planetesimals spiral into the planet.

These results show that temporary capture plays an important role in the capture and survival of planetesimals in the circumplanetary gas disk. 
Figure~\ref{fig:e0.5_tcap} shows the duration of temporary capture as a function of semi-major axis for captured planetesimals shown in Figures~\ref{fig:e0.5pro} and \ref{fig:e0.5ret}. 
In the case of prograde capture with relatively strong gas drag shown in Figure~\ref{fig:e0.5_tcap}(a), 
a significant number of capture occurs only in the case of short $\tau_{\rm dis}$ (green squares), and these prograde orbits experience a rather short period of temporary capture ($T_{\rm tc}/T_{\rm K}\lesssim 1$), meaning that they are captured by a single encounter with the disk. 
On the other hand, in the case of retrograde orbits with long $\tau_{\rm dis}$, planetesimals with a rather long period of temporary capture tend to survive (Figures~\ref{fig:e0.5_tcap}(b), (c)). 
In particular, when the gas drag is weak, remaining planetesimals are dominated by those experience long period of temporary capture (Figure~\ref{fig:e0.5_tcap}(c)).
Figure~\ref{fig:e0.5_tcap}(d) shows the plots of $T_{\rm tc}/\tau_{\rm dis}$ as a function of $a_{\rm p}/R_{\rm H}$ for those with $10^2 \lesssim \tau_{\rm dis}/ T_{\rm K} \lesssim 10^4$ out of the orbits shown in Figure~\ref{fig:e0.5_tcap}(b).
We can see that points for different $\tau_{\rm dis}$'s significantly overlap, 
which means that those planetesimals with  $T_{\rm tc} \sim \tau_{\rm dis}$ survive in the disk. 
Those with $T_{\rm tc} \ll \tau_{\rm dis}$ spiral into the planet because of the continued gas drag after capture, 
while those with $T_{\rm tc} \gg \tau_{\rm dis}$ are not captured because of the lack of energy dissipation from the short-lived gas disk.
Figure~\ref{fig:e0.5_tcap}(d) shows that planetesimals captured into orbits with large $a_{\rm p}$'s tend to have experienced longer duration of temporary capture.

\begin{figure}[htbp]
 \begin{center}
\plottwo{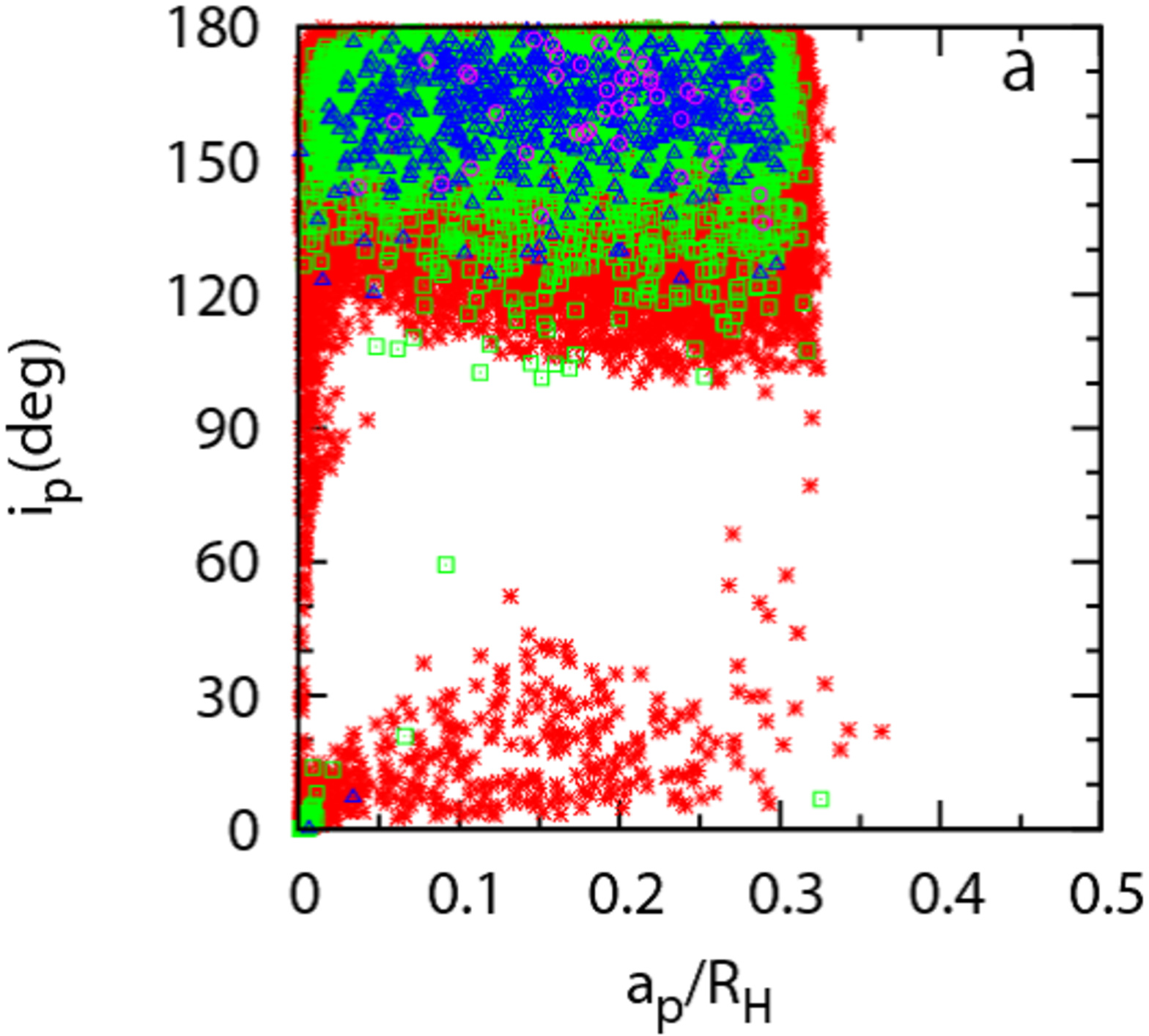}{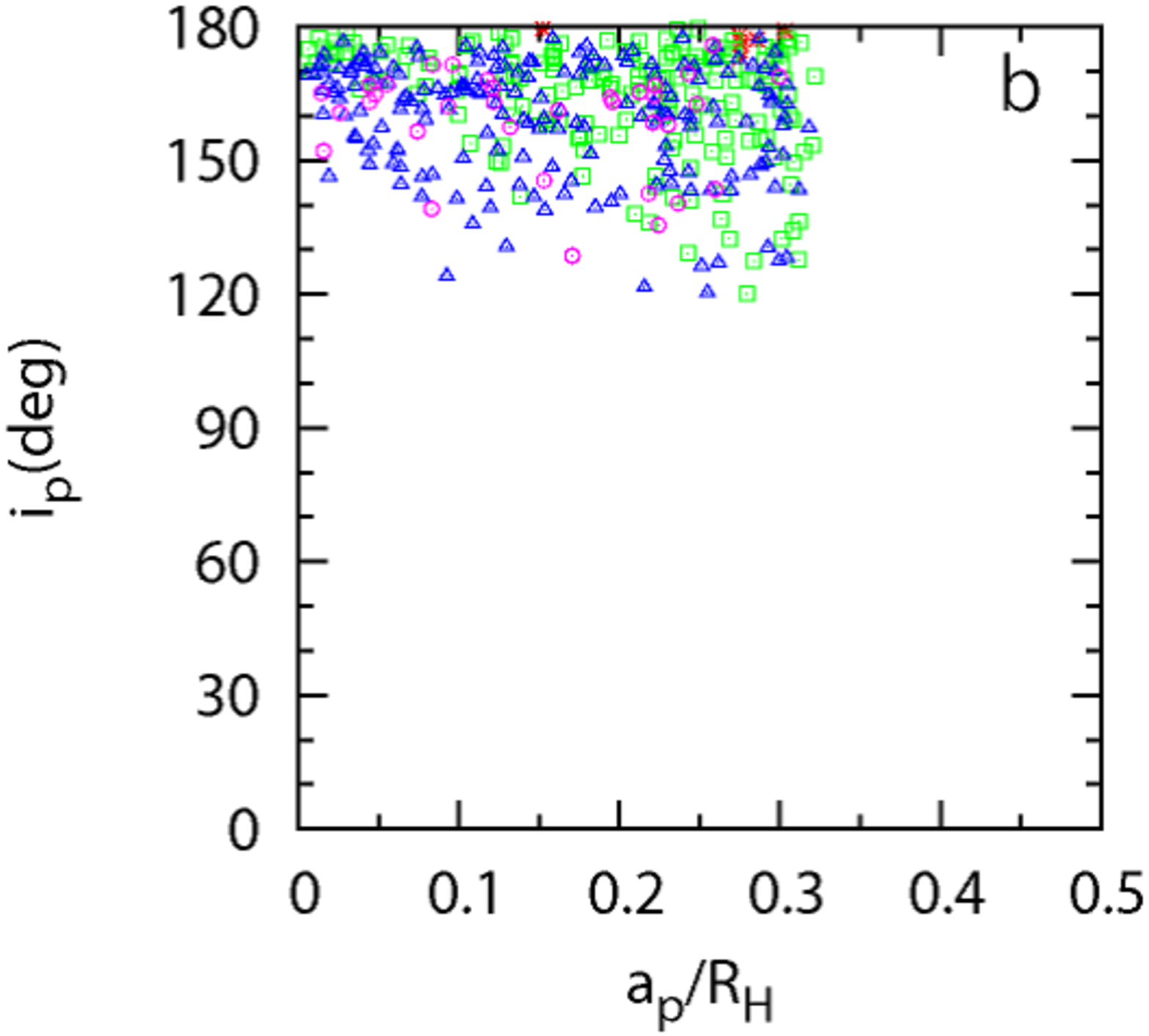}
 \end{center}
  \caption{Distribution of inclination ($i_{\rm p}$) of captured planetesimals at the end of simulation ($\langle e_{\rm H}^{2} \rangle^{1/2}=0.5$).
Panel (a) shows results with $\zeta_{\rm ini}=10^{-8}$, while Panel (b) shows those for $\zeta_{\rm ini}=10^{-10}$. 
Different marks represent different dispersal timescales.
Red asterisks, green squares, blue triangles and purple circles represent $\tau_{\rm dis}/T_{\rm K}=10, 10^{2}, 10^{3}$ and $10^{4}$, respectively.}
 \label{fig:e0.5ip}
\end{figure}

Figure~\ref{fig:e0.5ip} shows the plots of inclinations of planet-centered orbits as a function of semi-major axis at the end of simulation, for captured planetesimals shown by Figures~\ref{fig:e0.5pro} and \ref{fig:e0.5ret}. 
Panels (a) and (b) show results for different values of $\zeta_{\rm ini}$, and different marks represent the difference in the dispersal timescale ($\tau_{\rm dis}$). 
Highly inclined orbits in the prograde direction appear only when $\tau_{\rm dis}$ is short, because planetesimals cannot be captured into long-lived prograde orbits in the case of low velocity dispersion.
On the other hand, in the retrograde case, a significant number of planetesimals can be captured and survive even with large values of $\tau_{\rm dis}$.
When $\tau_{\rm dis}$ is short, captured planetesimals tend to retain highly off-plane orbits ($100^\circ \lesssim i_{\rm p} \lesssim 120^\circ$), while such orbits tend to disappear with increasing $\tau_{\rm dis}$.

\subsubsection{Case of Moderate Velocity Dispersion}
\begin{figure}[htbp]
 \begin{center}
\plottwo{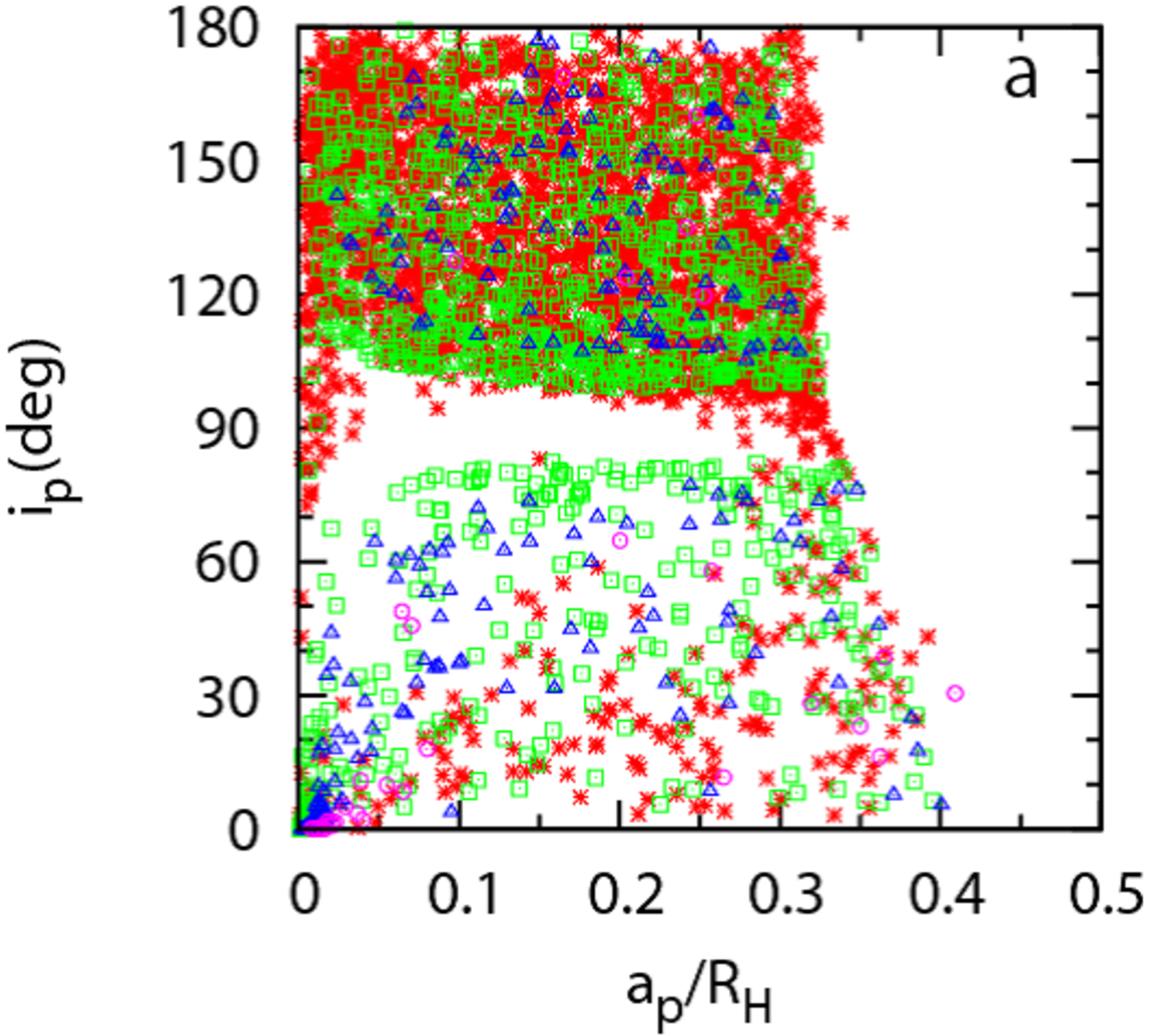}{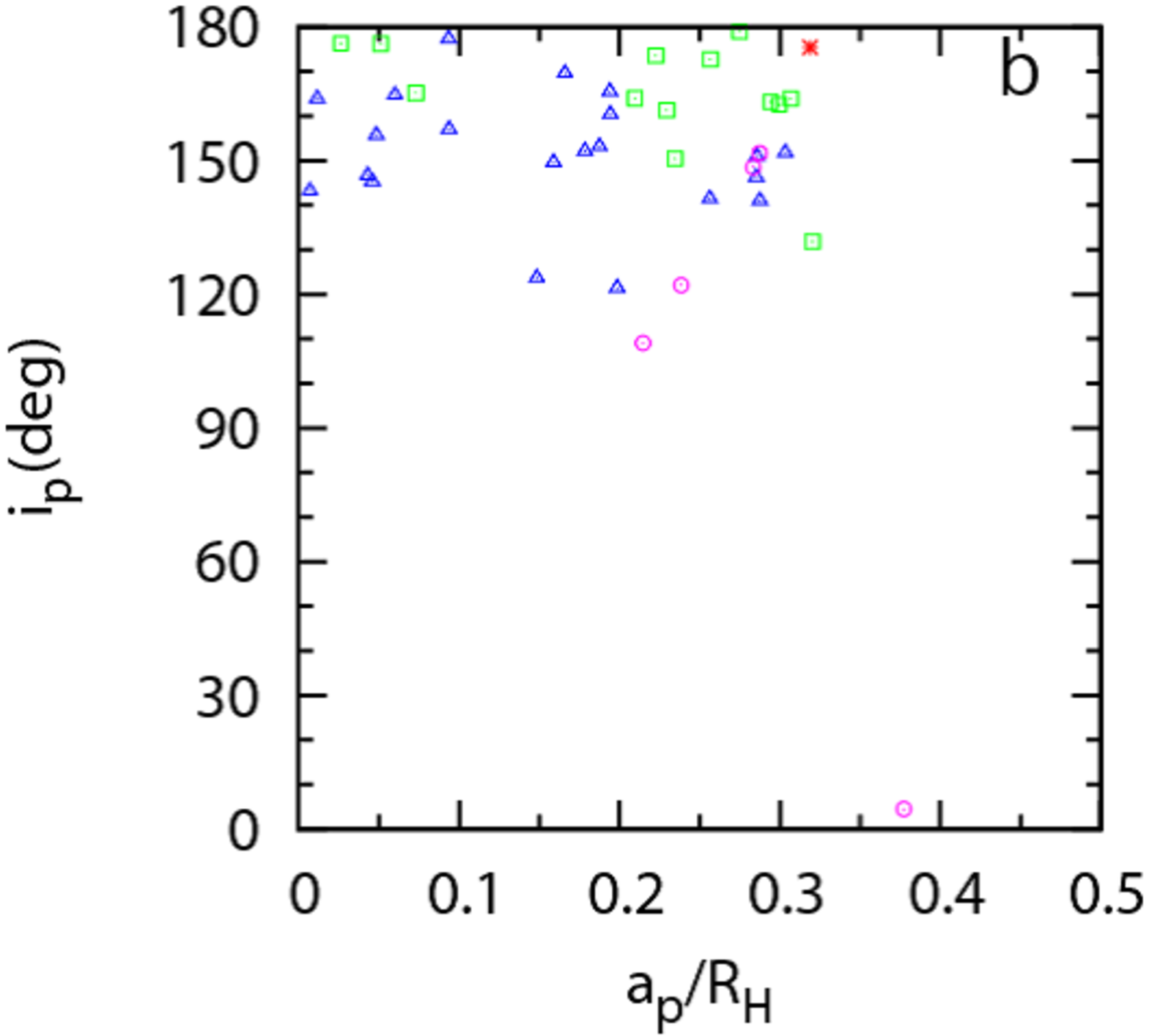}
 \end{center}
  \caption{Same as Figure~\ref{fig:e0.5ip}, but for the case of $\langle e_{\rm H}^{2} \rangle^{1/2}=2\langle i_{\rm H}^{2} \rangle^{1/2}=2$.}
 \label{fig:e2ip}
\end{figure}

Rates and duration of temporary capture sensitively depend on orbital elements of pre-capture heliocentric orbits \citep{S11,SO13}. Figure~\ref{fig:e2pro} shows the results for planetesimals captured into the prograde direction for the case with $\langle e_{\rm H}^{2} \rangle^{1/2}=2$ ($\zeta_{\rm ini}= 10^{-8}$).
In this case, owing to the contribution of the long-lived orbits with $e_{\rm H} \sim 3$, a significant number of planetesimals survive when $\tau_{\rm dis}$ is as long as $10^2-10^3$ years; there are a small number of surviving planetesimals even in the case with $\tau_{\rm dis} = 10^4$, where there are no surviving ones in the case of low-velocity dispersion (Figure~\ref{fig:e0.5pro}). 
Figure~\ref{fig:e2ret} shows orbital distribution for planetesimals captured in the retrograde direction ($\zeta_{\rm ini}=10^{-8}$ and $10^{-10}$). 
Long-lived retrograde orbits appear for planetesimals with energy in a certain range ($E \simeq 1.5 -2$), thus planetesimals with larger initial energy can be captured into such long-lived orbits if their energy is reduced by gas drag. 
In the case with rather strong gas drag ($\zeta_{\rm ini} = 10^{-8}$), capture via this kind of long-lived temporary capture is dominant. 
On the other hand, when the gas drag is too weak to reduce the energy to the above range, capture rates are rather small (Figure~\ref{fig:e2ret}(e) - (h)). 
Figure~\ref{fig:e2ip} shows the plots of inclinations of planet-centered orbits as a function of semi-major axis. 
The basic features are similar to the shear-dominated case (Figure~\ref{fig:e0.5ip}). 
However, we found that some of planetesimals captured into the prograde direction can have large inclinations when $\tau_{\rm dis}$ is rather long ($\tau_{\rm dis}=10^{2}-10^{3}T_{\rm K}$),
because planetesimals captured into long-lived prograde orbits can retain large inclinations for a long time (Figure~\ref{fig:orbit_pro3d}).

Both in the prograde and retrograde cases, we found that there is an upper limit of $a_{\rm p} \simeq (1/3)\times R_{\rm H}$ for captured planetesimals, as in the case of stronger gas drag \citep{SO16}.
This is consistent with our analytic estimate for the upper limit derived in Section~\ref{subsec:twoorb_ircap}.
Immediately after capture, the eccentricity of the planet-centered orbits of planetesimals is close to unity, thus its apocenter distance is about $2a_{\rm p}$. 
When the semi-major axis $a_{\rm p}$ has the above upper limit, the apocenter distance is smaller than $(2/3)\times R_{\rm H}$, 
i.e., smaller than the short axis of the planet's Hill sphere in the azimuthal direction, thus the captured planetesimals can stay within the Hill sphere.

\subsection{Capture Efficiency}

\begin{figure}[htbp]
\begin{center}
\plottwo{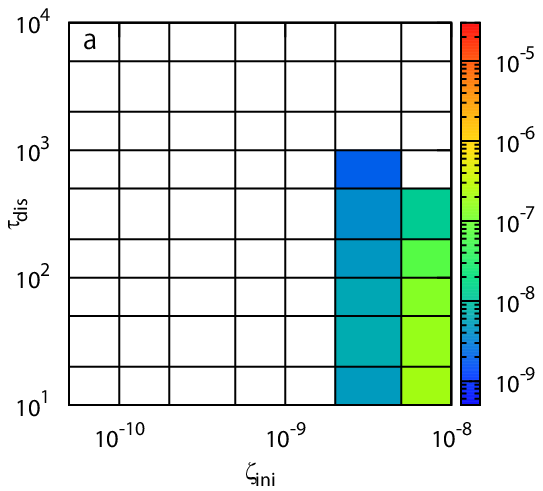}{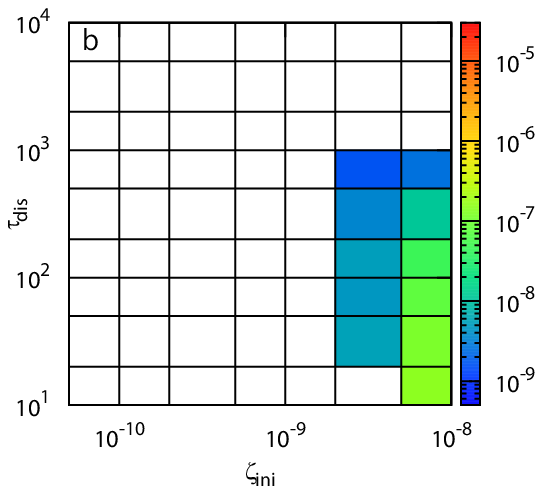}
\plottwo{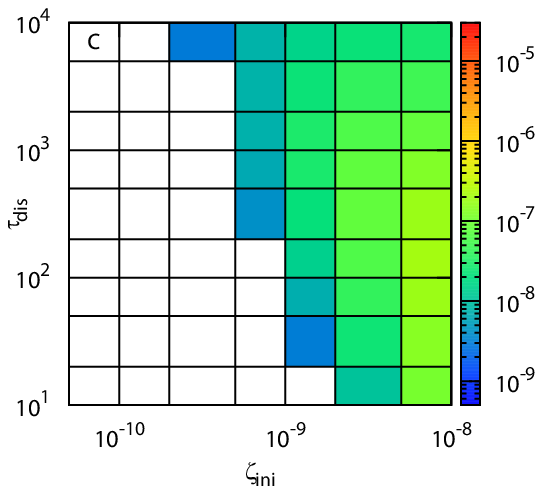}{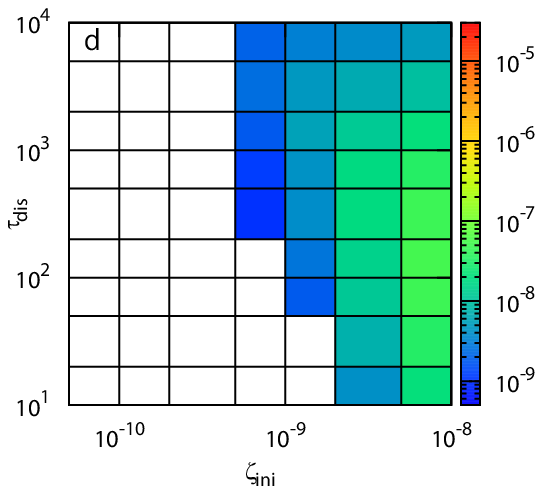}
\end{center}
 \caption{Efficiency of capture of planetesimals into planet-centered prograde orbits on the $\zeta_{\rm ini}$-$\tau_{\rm dis}$ plane.
$\langle e_{\rm H}^{2} \rangle^{1/2}=2\langle i_{\rm H}^{2} \rangle^{1/2}=0.2$(a), 0.7(b), 2(c), and 4(d).
The color in each area represents the average capture efficiency for the four vertices of the area. 
}
 \label{fig:cont_pro}
\end{figure}

In the simulations presented in Section 5.2, we integrated about $10^9$ orbits for a given set of parameters (i.e., $\langle e_{\rm H}^{2} \rangle^{1/2} $, $\zeta_{\rm ini}$, $\tau_{\rm dis}$). 
Thus, by dividing the number of surviving captured planetesimals by this number, we can estimate capture efficiency. 
Figure~\ref{fig:cont_pro} shows the plots of the capture efficiency in the prograde case on the $\tau_{\rm dis}-\zeta_{\rm ini}$ plane for four different values of the velocity dispersion.
In the case of the shear-dominated regime ($\langle e_{\rm H}^{2} \rangle^{1/2}=0.2$ and $0.7$; 
Figure~\ref{fig:cont_pro}(a)(b)), we find that capture takes place in a rather narrow region with strong gas drag and short dispersal timescale ($\zeta_{\rm ini} \simeq 5 \times10^{-9}-10^{-8}$ and $\tau_{\rm dis} \simeq10-5 \times10^{2} T_{\rm K}$).
Long-lived capture orbits do not play a role because planetesimals' initial velocity dispersion is small and their energy is large. 
On the other hand, in the case of the dispersion-dominated regime (Figures~\ref{fig:cont_pro}(c) and (d); 
$\langle e_{\rm H}^{2} \rangle^{1/2}=2$ and 4), 
planetesimals are captured even with weak gas drag and/or slow disk dispersal ($\tau_{\rm dis}\gtrsim10^{3}T_{\rm K}$), 
because they become captured into long-lived prograde orbits.
However, if the gas drag is too weak (i.e., $\zeta_{\rm ini}\lesssim5\times10^{-10}$), capture does not take place.

\begin{figure}[htbp]
 \begin{center}
\plottwo{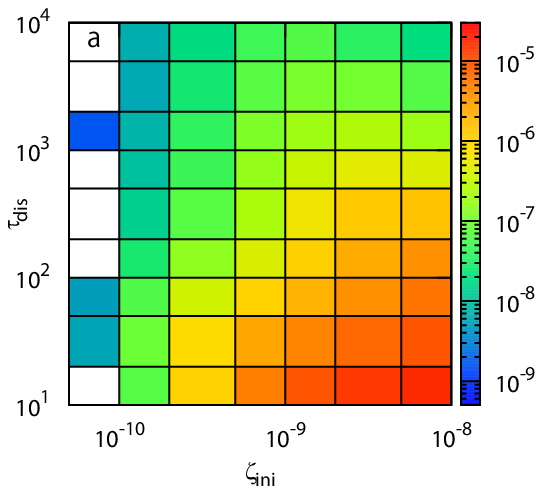}{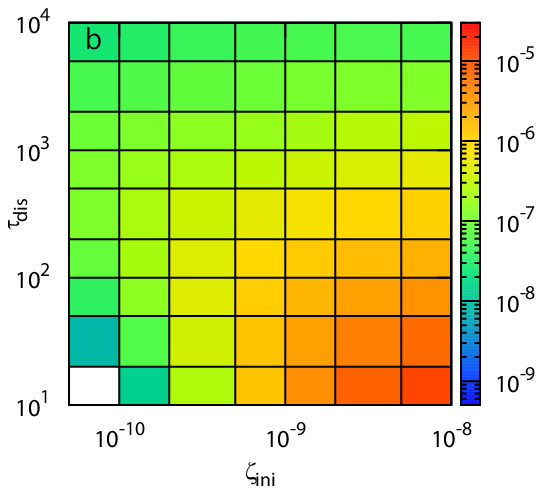}
\plottwo{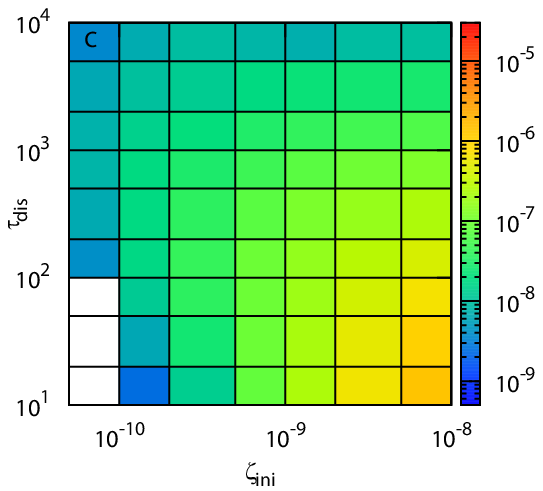}{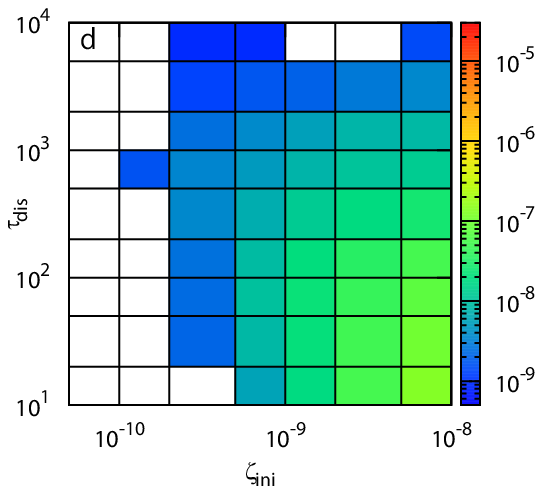}
 \end{center}
 \caption{Same as Figure~\ref{fig:cont_pro}, but for the case of capture in the retrograde direction.}
 \label{fig:cont_ret}
\end{figure}

Figure~\ref{fig:cont_ret} shows the efficiency of capture into retrograde orbits. 
Capture takes place for a wider range of $\tau_{\rm dis}$ and $\zeta_{\rm ini}$ compared to the prograde case, and the efficiencies are also larger.
When planetesimals have low initial velocity dispersions, their energy is too high to become captured into long-lived orbits; 
they can be captured if their energy is reduced to the value corresponding to the long-lived capture orbits and then the gas disk is dissipated in a short timescale. 
Thus, the capture efficiency in this case has a peak in the narrow region with large $\zeta_{\rm ini}$ and small $\tau_{\rm dis}$.
When the velocity dispersion is large, capture efficiency becomes lower, as we mentioned in the previous sections.

These numerical results show that strong initial gas drag and quick dispersal of the circumplanetary gas disk facilitates capture and survival of planetesimals, 
and long-lived capture orbits would play an important role for permanent capture depending on parameters.
Capture in the prograde (retrograde) direction more easily takes place with planetesimals' 
initial random velocities in the dispersion-dominated (shear-dominated) regime.

\section{SUMMARY AND DISCUSSION}
\label{sec:sum}

In the present work, we performed three-body orbital integration for the capture of planetesimals and their subsequent orbital evolution under gas drag.
We found that capture process of planetesimals depends on the strength of gas drag. 
In the case of relatively strong gas drag, capture typically takes place in a single encounter with the planet \citep{F13}. 
On the other hand, permanent capture via temporary capture phase is dominant when gas drag is weak. 
Temporarily captured planetesimals interact with the circumplanetary disk many times, thus even weak gas drag can lead to permanent capture. 
We found that there are certain types of capture orbits in both prograde and retrograde directions about the planet that allow survival of captured planetesimals in the circumplanetary disk for a long period of time under weak gas drag. 
Long-lived prograde capture occurs only for a limited range of eccentricity ($e_{\rm H}\simeq3$) with low energy.
This behavior is similar to the case of the gas-free environment \citep{S11}.
On the other hand, long-lived retrograde capture occurs at $e_{\rm H}\simeq0.7$.
This types of long-lived orbits were not found in the the gas-free environment, and gas drag assisted capture into such orbits. 

We also examined the distribution of captured planetesimals after the dispersal of circumplanetary disk, taking account of gradual dispersal of the circumplanetary disk.
One notable feature common in the prograde and retrograde cases is that there seems to be an upper limit for the values of the semi-major axes of captured planetesimals at about $a_{\rm p}\simeq R_{\rm H}/3$, which can be explained by a simple analytic consideration \citep{SO16}. 
We found that final distribution of planet-centered orbits of captured planetesimals depends on the strength of gas drag and timescale of disk dispersal. 
When the gas drag is strong and the disk dispersal takes place in a short timescale, planetesimals are captured into orbits with small $a_{\rm p}$ by a single encounter with the planet.
Planetesimals that experience multiple encounters with the phase of temporary capture settle into orbits with large $a_{\rm p}$.
On the other hand, if the timescale of disk dispersal is rather long, the duration of temporary capture becomes long.
When gas drag is weak and the timescale for the disk dispersal is too short, no planetesimals survive, because the capture via the long phase of temporary capture does not work in such a case. 
In this case, permanent capture takes place when the timescale of disk dispersal is long.

We have assumed that the radial distribution of planetesimals in the protoplanetary disk is uniform. 
However, it is not clear if this assumption is applicable to the late stage of the giant planet's formation. 
Source regions for planetesimals captured into long-lived prograde orbits are $b_{\rm H}\simeq4-5.5$, while $b_{\rm H}\simeq2-3.5$ for long-lived retrograde orbits (see also \citet{S11}). 
Thus, the effect of the gap in the planetesimal disk on capture efficiency is very small when the gap width is smaller than $2R_{\rm H}$.
However, when the width of the gap is further increased,  the capture efficiency decreases significantly, because most of planetesimals approaching the planet's Hill sphere are removed \citep{F13}.
\citet{DP15} examined orbital evolution of planetesimals in the vicinity of Jupiter and their distribution in the circumjovian disk after permanent capture.
Their results show that the width of the gap formed by Jupiter is $\sim3R_{\rm H}$.
Thus, the efficiency of capture into long-lived retrograde orbits would decrease due to the effect of the gap, while the effect on long -lived prograde capture is expected to be small.
 
In the present work, we also assumed that the mass of planetesimals is unchanged during orbital integration.
However, it may be reduced by ablation during the passage through the circumplanetary disk, and such mass loss may affect capture rates of planetesimals and their orbital evolution \citep{F13, DP15}.
Thus, we examined the influence of ablation on long-lived capture orbits, using three-body orbital integration including the effect of mass loss due to ablation (see \citet{F13} for the methods of calculation).
We found that the effect of ablation on temporarily captured planetesimals is insignificant,
because planetesimals do not pass through the dense part of the circumplanetary disk during temporary capture.
After permanently captured, the mass of planetesimals is reduced by ablation, depending on the orbital elements of their planet-centered orbits.
As mentioned above, planetesimal have large eccentricities and semi-major axes of planet-centered orbits immediately after permanent capture, and the mass loss mostly occurs during a short period of time at their pericenter passage.
Consequently, the effect of ablation on captured planetesimals with large semi-major axes ($a_{\rm p}\gtrsim0.1R_{\rm H}$) is expected to be insignificant.
On the other hand, if semimajor axes become small enough by gas drag during orbital evolution,
planetesimals constantly pass through the dense part of the circumplanetary disk.
Thus, the mass of planetesimals that have spiraled into orbits in the  vicinity of the planet would be reduced significantly by ablation.

\begin{figure}[htbp]
 \begin{center}
  \includegraphics[width=165mm]{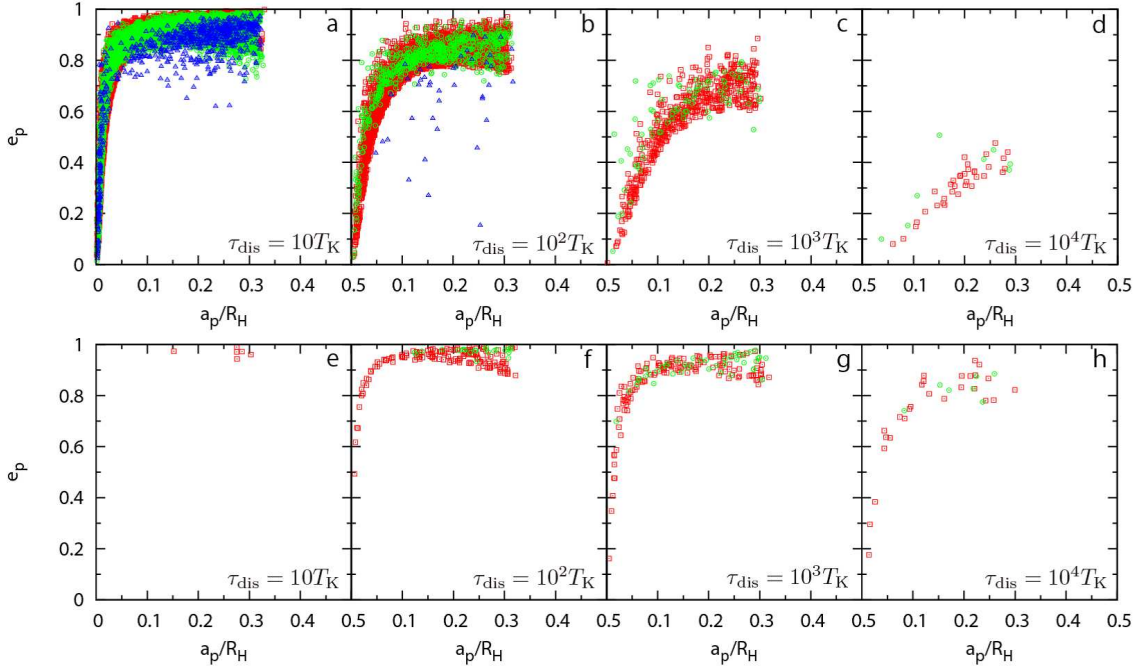}
 \end{center}
  \caption{Distribution of orbital elements of planetesimals captured into retrograde orbits at the end of simulation ($\langle e_{\rm H}^{2} \rangle^{1/2}=0.5$).
Panels (a)-(d) are the results with $\zeta_{\rm ini}=10^{-8}$, while Panels (e)-(h) are those for $\zeta_{\rm ini}=10^{-10}$. 
Panels (a), (b), (c) and (d) (Panels (e), (f), (g) and (h)) show the results for $\tau_{\rm dis}=10, 10^{2}$, $10^{3}$ and $10^{4}T_{\rm K}$, respectively. 
Different marks represent inclinations of planet-centered orbits.
Triangles, circles and squares represent $90^\circ\leq i_{\rm p}<120^\circ$, $120^\circ\leq i_{\rm p}<150^\circ$ and $150^\circ\leq i_{\rm p}<180^\circ$, respectively.}
 \label{fig:e0.5ret}
\end{figure}

\begin{figure}[htbp]
\begin{center}
\plottwo{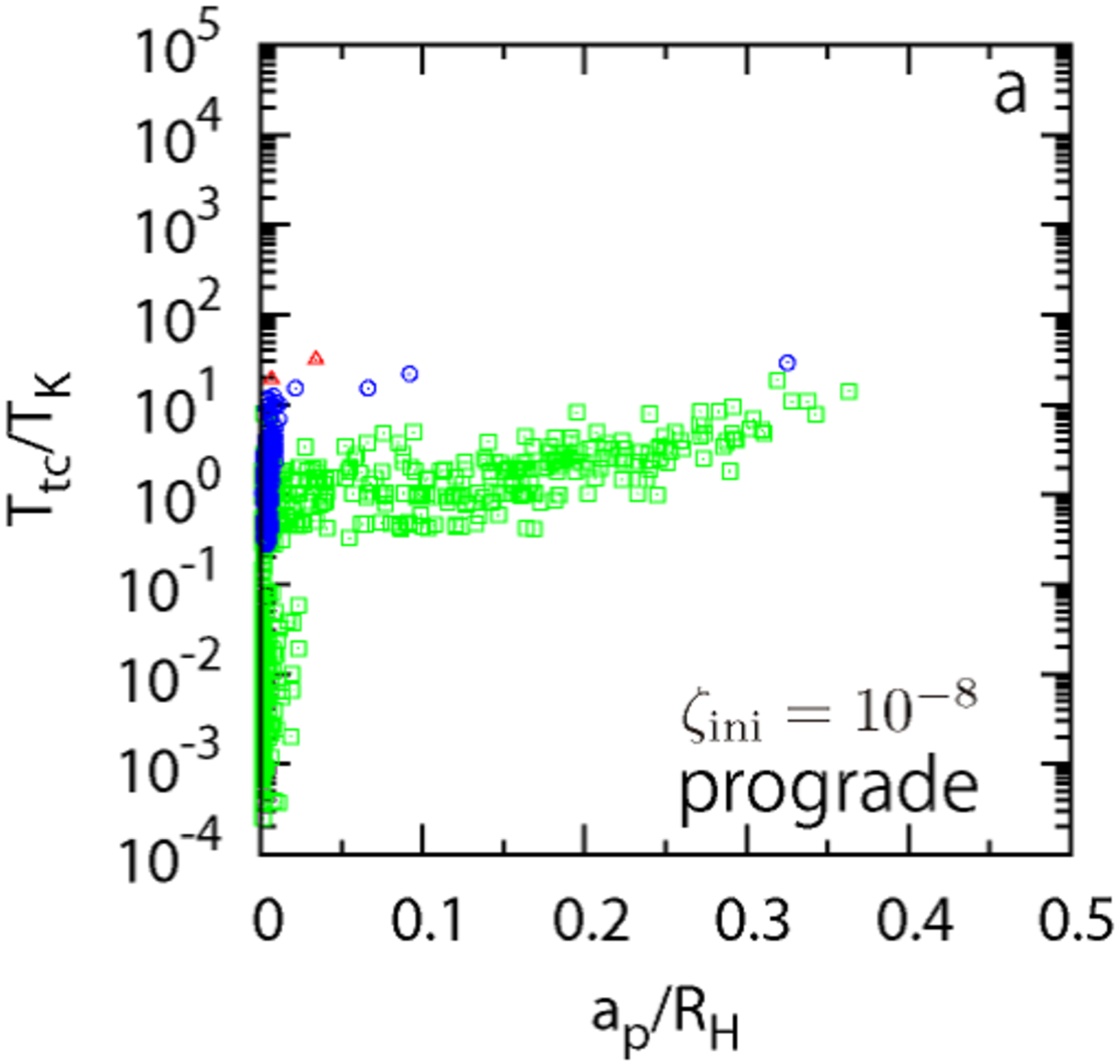}{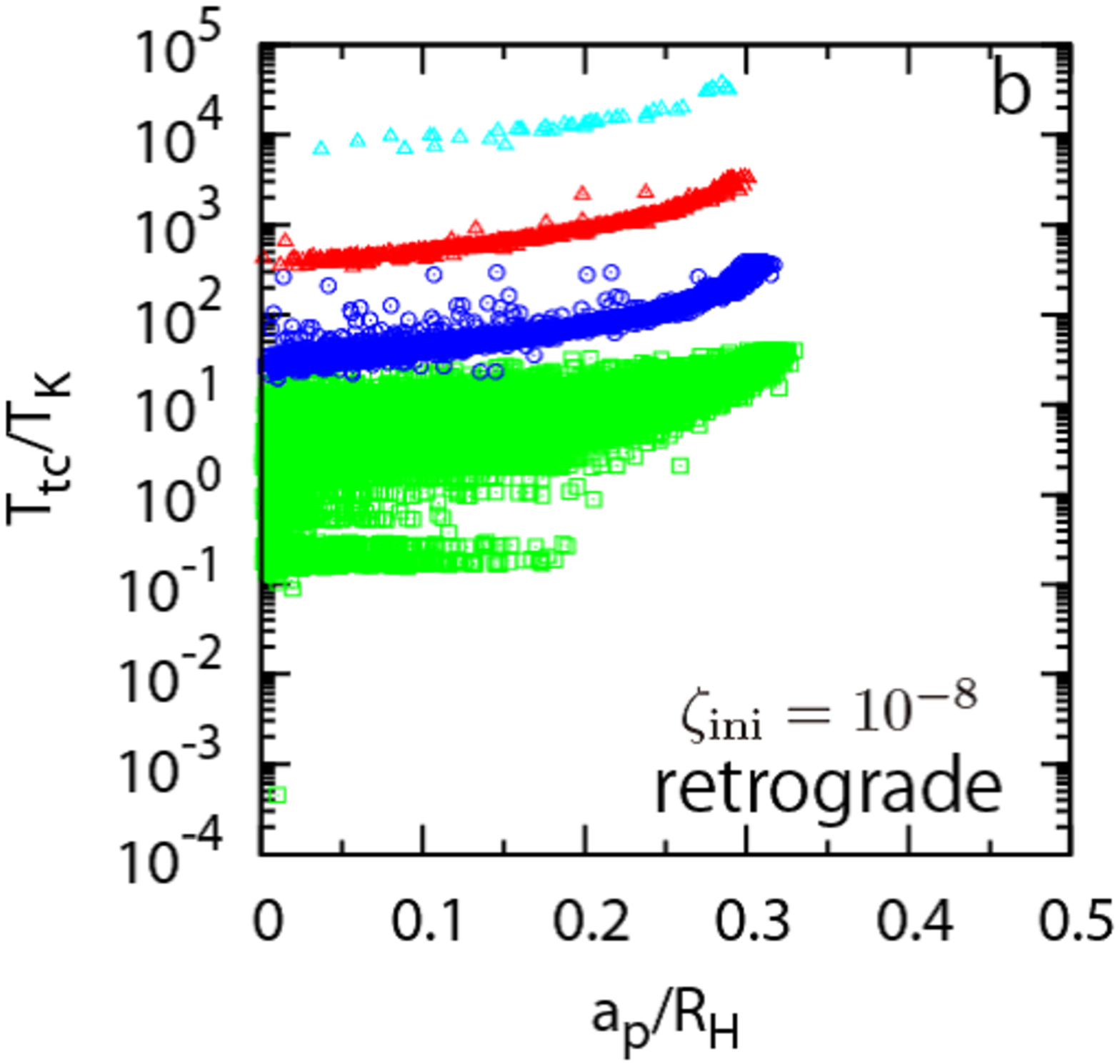}
\plottwo{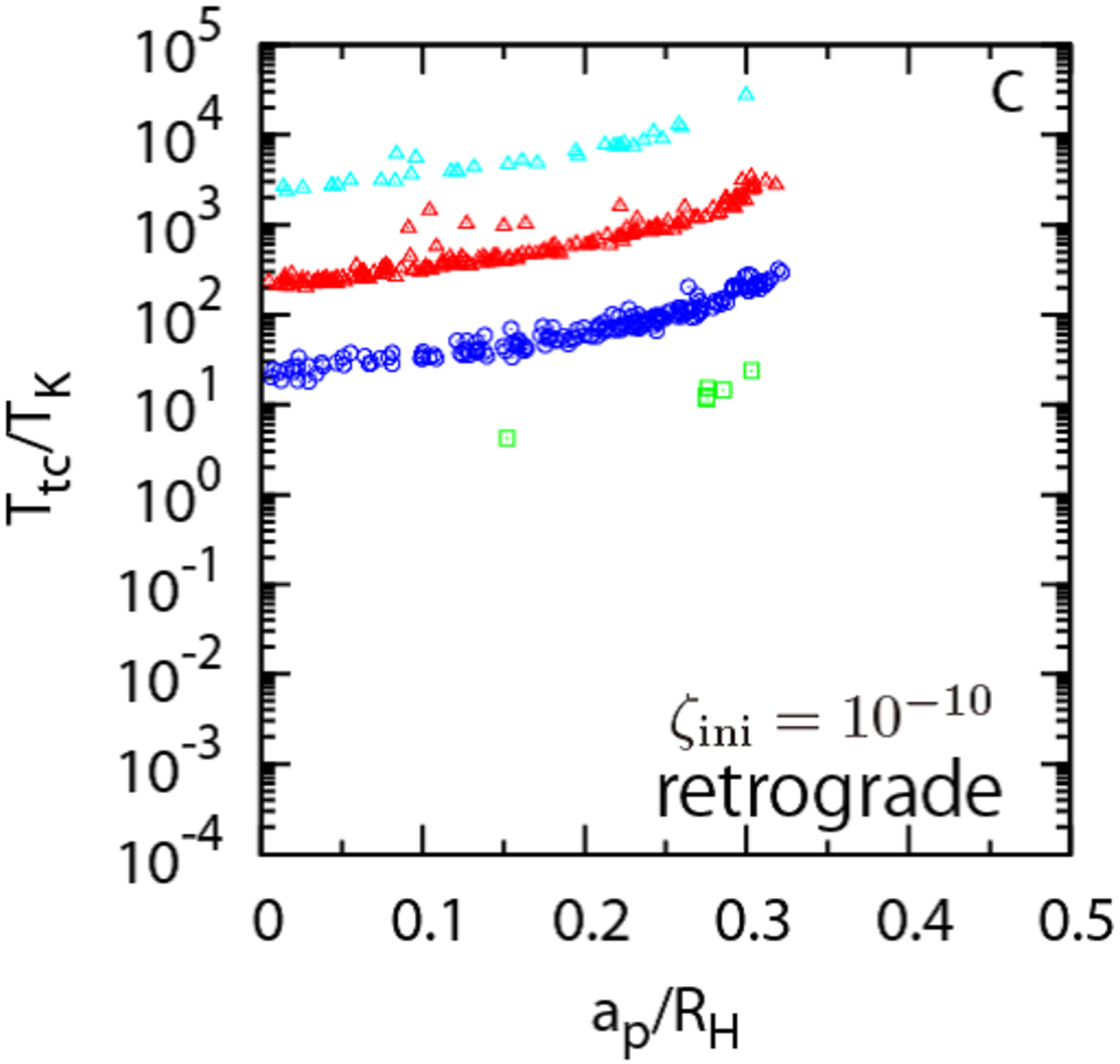}{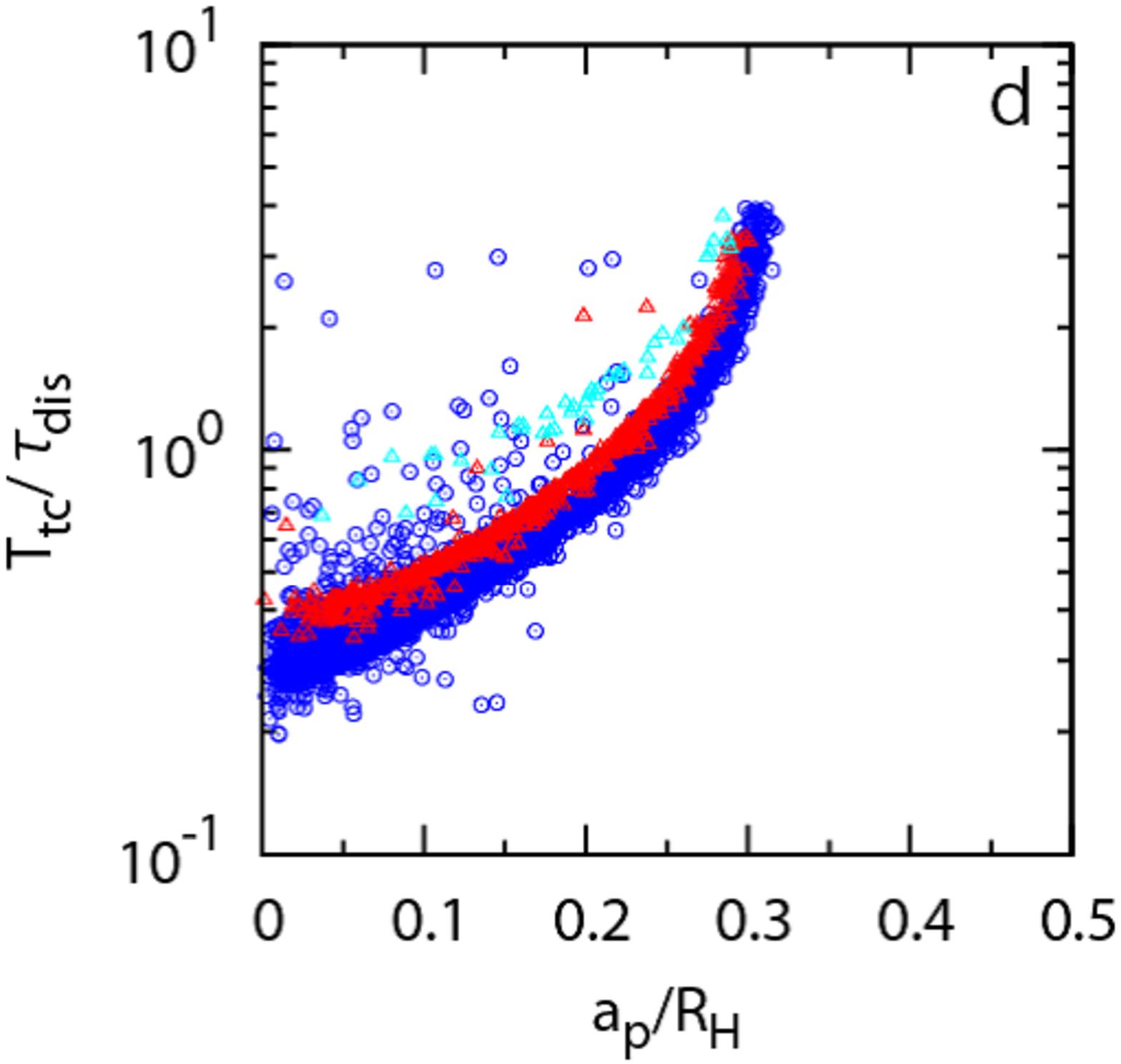}
\end{center}
  \caption{Relationship between the semi-major axis and the duration of temporary capture.
(a) Duration of temporary capture for the case shown in Figure~\ref{fig:e0.5pro}.
Green, blue, red and cyan marks represent $\tau_{\rm dis}/T_{\rm K}=10, 10^{2}, 10^{3}$ and $10^{4}$, respectively.
(b) Results for the cases shown in Figures~\ref{fig:e0.5ret}(a)-(d). 
(c) Cases shown in Figure~\ref{fig:e0.5ret}(e)-(h). 
(d) Normalized timescale $T_{\rm tc}/\tau_{\rm dis}$ for the case shown in Figure~\ref{fig:e0.5_tcap}(b). 
Blue, red and cyan marks represent $\tau_{\rm dis}/T_{\rm K}=10^{2}, 10^{3}$ and $10^{4}$, respectively.}
 \label{fig:e0.5_tcap}
\end{figure}

\begin{figure}[htbp]
 \begin{center}
  \includegraphics[width=165mm]{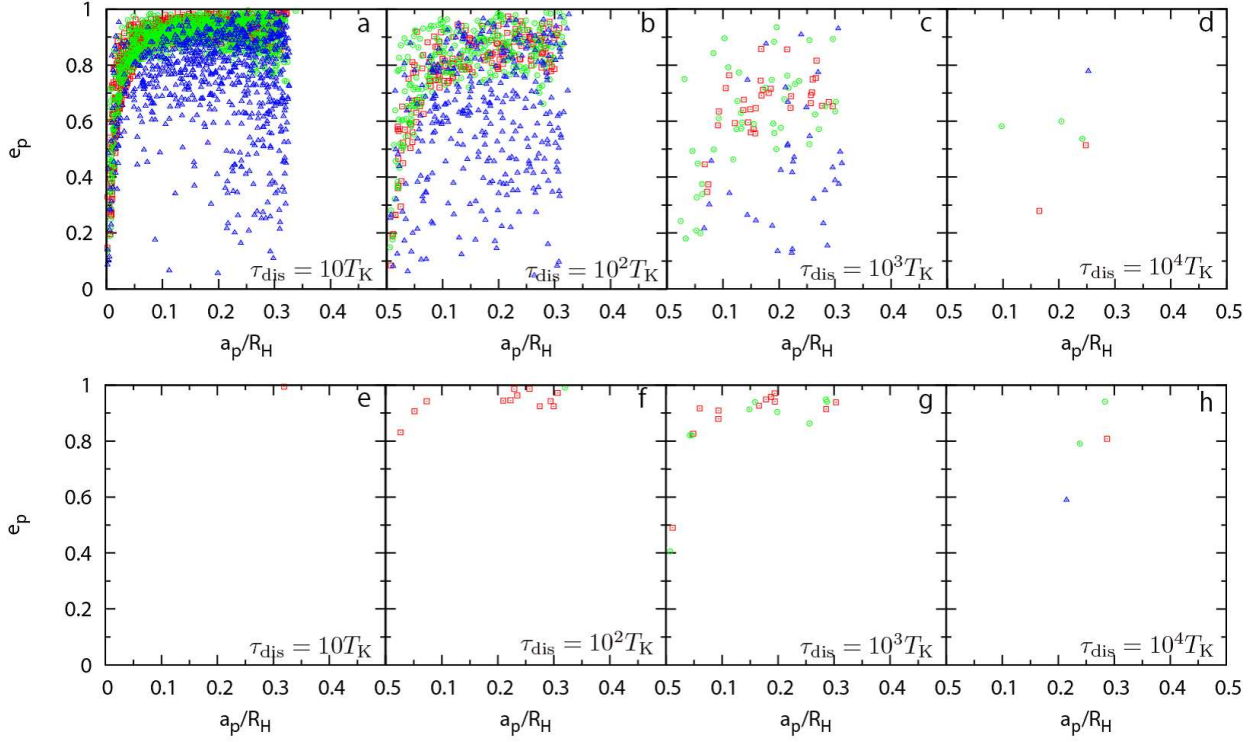}
 \end{center}
  \caption{Same as Figure~\ref{fig:e0.5ret}, but for the case of $\langle e_{\rm H}^{2} \rangle^{1/2}=2\langle i_{\rm H}^{2} \rangle^{1/2}=2$.
Panels (a)-(d) show results for $\zeta_{\rm ini}=10^{-8}$, while Panels (e)-(h) are those for $\zeta_{\rm ini}=10^{-10}$.
} 
 \label{fig:e2ret}
\end{figure}

Our numerical results provide some insights into capture process of irregular satellites.
Irregular satellites of the giant planets in the Solar System have values of $a/R_{\rm H}$ between 0.1 and 0.5 \citep{JH07, B10}, 
with the prograde ones having smaller semi-major axes ($0.1-0.3R_{\rm H}$) than the retrograde ones ($0.2-0.5R_{\rm H}$). 
Their eccentricities range $0.1 <e_{\rm p}< 0.7$, and the region of their inclinations between $\sim60^\circ$ and $\sim120^\circ$ contains no satellites due to the Kozai resonance \citep{N03}.
Our simulations show that surviving planetesimals would have relatively small eccentricities ($\lesssim0.4$) in the case of large $\zeta_{\rm ini}$ and long $\tau_{\rm dis}$ (Figure~\ref{fig:e0.5ret}(d)).
However, it is difficult to explain capture of irregular satellites with large semi-major axes by gas drag alone, because captured planetesimals have $a_{\rm p}\lesssim R_{\rm H}/3$.
The difference in semi-major axes cannot be explained by subsequent evolution such as tidal evolution, because tidal force from the planet is too weak. 
Therefore, other capture models such as those based on purely gravitational interactions seem to be required for the capture of such irregular satellites.

\begin{figure}[htbp]
 \begin{center}
  \includegraphics[width=165mm]{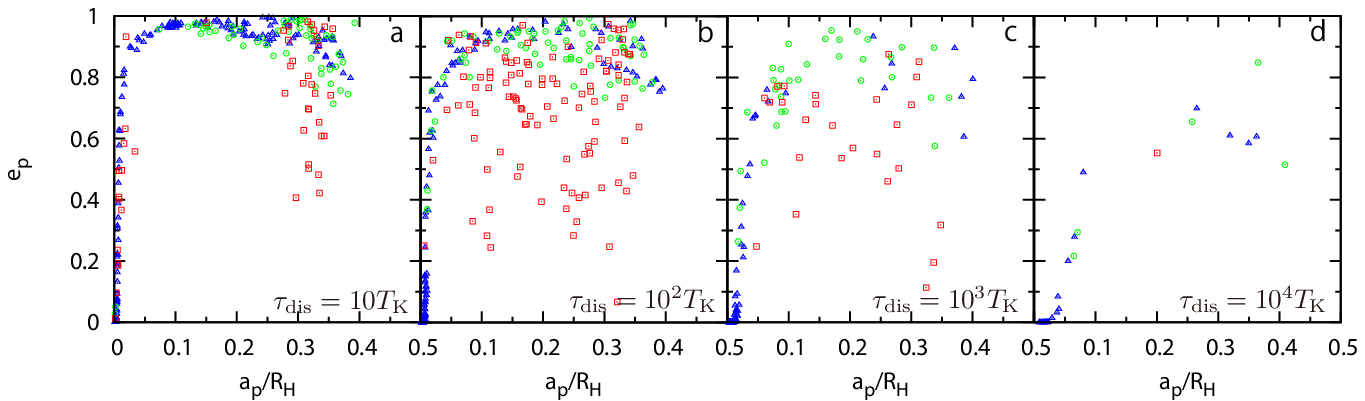}
 \end{center}
  \caption{Same as Figure~\ref{fig:e0.5pro}, but for the case of $\langle e_{\rm H}^{2} \rangle^{1/2}=2\langle i_{\rm H}^{2} \rangle^{1/2}=2$.}
 \label{fig:e2pro}
\end{figure}

Recent studies show that capture of irregular satellites by three-body interaction seems to be promising \citep{N07, V08, P10, G11, G13, N14}.
For example, \citet{N07, N14} examined capture of irregular satellites by three-body interaction among two planets and a neighboring planetesimal during close encounter between the planets, and showed that a sufficient number of planetesimals can be captured to explain observed irregular satellites.  
\citet{N14} showed that some of Jupiter's irregular satellites with 0.05AU$<a_{\rm p}<$0.1AU (0.14$<a_{\rm p}/R_{\rm H, J}<$0.29, where $R_{\rm H, J}$ is Jupiter's Hill radius) can survive perturbation due to close encounters with other planets. 
Thus, a part of planetesimals captured by gas drag into orbits with small semi-major axes would survive the close encounter.  
Moreover, in the above model, it is difficult to explain capture of Jupiter's largest irregular satellite Himalia because of the low capture efficiency \citep{N14}. 
As mentioned above, if initial energy of planetesimals is very low, they are likely to be captured into prograde orbits as Himalia.
Our results (Figures~\ref{fig:e2pro}(a), (b)) show that semi-major axes of captured planetesimals are consistent with the semi-major axis of Himalia if $\tau_{\rm dis}\leq10^{2}T_{\rm K}$; such a short dispersal timescale would be possible if Jupiter opened a complete gap in the protoplanetary disk and  the infall onto the circumplanetary disk was cutoff quickly.
However,  eccentricities and inclinations of captured planetesimals in such a case are larger than that of Himalia.
Therefore, other mechanisms seem to be needed to fully explain its origin.

Our results suggest that giant planets would have many planetesimals captured by gas drag immediately after the dispersal of the circumplanetary disk.
If collisional grinding of such captured planetesimals occurs around exoplanets and produces a sufficient amount of dusts, they may be observable and provide us with important constraints on the evolution of satellite system of exoplanets \citep{KW11}.
Captured planetesimals would also contribute to the formation of craters on regular satellites.
Further studies are needed to clarify the effects of captured planetesimals on the evolution of satellite systems.

\acknowledgments
This work was supported by JSPS Grants-in-Aid for JSPS Fellows (12J01826) and Scientific Research B (22340125 and 15H03716).
Part of numerical calculations were performed using computer systems at the National Astronomical Observatory of Japan.

\clearpage

\clearpage

\end{document}